# Simply Amusing Algebra and Analysis or Electromagnetic and Gravitational Fields in the Single System of Equations?

D. M. Volokitin

The Special Theory of Relativity is based on one assumption only – the finiteness of speed of distribution of material objects. The General Theory of Relativity is based on Special one. One of the solutions of the equations of General Theory, the solution of Schwarzschild, is stationary, i.e. such at which influence of a source of a field on a trial particle does not depend on time, and hence spreads instantly. Isn't it?
Something here is not correct …

*One of the questions that Alice could have asked Charles Dodgson 137 years later.*

## Introduction.

I warn in advance: I am not a professional physicist-theorist. More precisely, I can not consider myself completely as those though at one time I graduated from theoretical physics faculty of physical department of Moscow State University.

Now, when I thus have forgiven to myself and have secured myself against charges in megalomania, I hope that I can count on your indulgence at reading the given article, and I can state my ideas more or less freely.

The basic idea of this article consists in the following. The physical space-time with all evidence is four-dimensional, at least, the experimental facts denying this statement are unknown to me. Therefore, if to be guided by "Occam's razor" it is not necessary to increase dimension of physical space-time in theoretical researches, while all possible ways of construction of the adequate theory of four-dimensional space are not tried. Here rather modest attempt (with obviously not complete results) to plan one of possible directions of construction of such theory is undertaken.

In my opinion, in four-dimensional space-time the algebra, corresponding it pseudoeuclidivity, is not constructed till now. In particular, I do not agree with Yu. S. Vladimirov, whose special seminar on gravitation I've attended, that algebras of quaternions and octaves "are the hypercomplex systems integrally comprising the signature $(+--...)$" [5, p.243]. On the contrary, these algebras as, however, algebra of complex numbers, contain the signature $(++...)$ because multiplication of an element by conjugated one gives in coordinate record especially euclidean expression $(x^0)^2+(x^1)^2+...$.

Further Vladimirov writes: "By means of quaternion algebra it is possible to construct good model of 4-dimensional space-time attitudes … Cleanly imaginary quaternions give model of 3-dimensional classical space $R^3$; algebraic product of two imaginary quaternions gives as the real part scalar, and as imaginary - vector product of corresponding vectors in $R^3$". However,

Vladimirov writes nothing about algebraic product of full quaternions, some of components of which, as it is known, do not have analogues in vector algebra.

Equally Hawking's idea about imaginary time in cosmological models has the same lack.

It seems to me, that the matter is that such structures (as just mentioned) do not make algebra. I.e. in a sense do not satisfy aesthetic criterion of correctness of the theory.

In this article the attempt to construct algebra of four-dimensional space-time is undertaken. Let's start with two requirements. The first is already mentioned pseudoeuclidivity. The second is an attempt to keep though any of "good" properties of algebras (associativity, commutability, presence of unity and so forth). In this case it is associativity. These two requirements are enough for unequivocal (to within several arbitrary real constants) construction of algebra.

The last stage of ideology of the present article is construction of elements of the theory of four-dimensional functions of four-dimensional variable.

And, at last, I wish to emphasize, that I do not insist at all on adequacy of the mathematical formalism stated in the present article to a real physical picture. Principles (of algebra construction) can be another. But it seems to me, that the idea of construction of algebra and the corresponding analysis in real space-time has the right to exist. This statement can be emphasized even by the fact of absence of central singularity in the solution of spherically-symmetric problem.

In conclusion of the Introduction I wish to note, that I tried to refrain in paragraphs 1-7 from any comments from the point of view of physicist. If there are results, about what to do inferences, let everyone reading will make them himself. My inferences are in the Conclusion. It is not obligatory to read it.

## 1. Algebra in Flat Space.

We shall consider four-dimensional linear space with basis $\{\vec{e}_i\}$. Let us enter in this space algebra by means of the instruction of rules of multiplication of basic vectors $\vec{e}_i$.

Let's name a vector $\vec{r}' = x^0\vec{e}_0 - x^1\vec{e}_1 - x^2\vec{e}_2 - x^3\vec{e}_3$ conjugated to a vector $\vec{r} = x^0\vec{e}_0 + x^1\vec{e}_1 + x^2\vec{e}_2 + x^3\vec{e}_3$.

Let's demand the product of any vector and its conjugated one to be equal to $((x^0)^2 - (x^1)^2 - (x^2)^2 - (x^3)^2)\vec{f}_0$ with a scalar multiplier before vector $\vec{f}_0$ looking like pseudo-norm of four-dimensional pseudoeuclidean space.

So, let

$$\begin{aligned}\vec{r}\vec{r}' &= (x^0\vec{e}_0 + x^1\vec{e}_1 + x^2\vec{e}_2 + x^3\vec{e}_3)(x^0\vec{e}_0 - x^1\vec{e}_1 - x^2\vec{e}_2 - x^3\vec{e}_3) = \\ &= (x^0)^2\vec{e}_0\vec{e}_0 + x^0x^1\vec{e}_1\vec{e}_0 + x^0x^2\vec{e}_2\vec{e}_0 + x^0x^3\vec{e}_3\vec{e}_0 - \\ &- x^0x^1\vec{e}_0\vec{e}_1 - (x^1)^2\vec{e}_1\vec{e}_1 - x^1x^2\vec{e}_2\vec{e}_1 - x^1x^3\vec{e}_3\vec{e}_1 - \\ &- x^0x^2\vec{e}_0\vec{e}_2 - x^1x^2\vec{e}_1\vec{e}_2 - (x^2)^2\vec{e}_2\vec{e}_2 - x^2x^3\vec{e}_3\vec{e}_2 - \\ &- x^0x^3\vec{e}_0\vec{e}_3 - x^1x^3\vec{e}_1\vec{e}_3 - x^2x^3\vec{e}_2\vec{e}_3 - (x^3)^2\vec{e}_3\vec{e}_3 = \\ &= ((x^0)^2 - (x^1)^2 - (x^2)^2 - (x^3)^2)\vec{f}_0\end{aligned} \quad . \tag{1.1}$$

Then, obviously, parities

$$
\begin{aligned}
&\vec{e}_0\vec{e}_0 = \vec{e}_1\vec{e}_1 = \vec{e}_2\vec{e}_2 = \vec{e}_3\vec{e}_3 = \vec{f}_0 = p_{00}\vec{e}_0 + p_{01}\vec{e}_1 + p_{02}\vec{e}_2 + p_{03}\vec{e}_3 \\
&\vec{e}_1\vec{e}_0 = \vec{e}_0\vec{e}_1 = \vec{f}_1 = p_{10}\vec{e}_0 + p_{11}\vec{e}_1 + p_{12}\vec{e}_2 + p_{13}\vec{e}_3 \\
&\vec{e}_2\vec{e}_0 = \vec{e}_0\vec{e}_2 = \vec{f}_2 = p_{20}\vec{e}_0 + p_{21}\vec{e}_1 + p_{22}\vec{e}_2 + p_{23}\vec{e}_3 \\
&\vec{e}_3\vec{e}_0 = \vec{e}_0\vec{e}_3 = \vec{f}_3 = p_{30}\vec{e}_0 + p_{31}\vec{e}_1 + p_{32}\vec{e}_2 + p_{33}\vec{e}_3 \\
&\vec{e}_1\vec{e}_2 = -\vec{e}_2\vec{e}_1 = \vec{g}_1 = q_{10}\vec{e}_0 + q_{11}\vec{e}_1 + q_{12}\vec{e}_2 + q_{13}\vec{e}_3 \\
&\vec{e}_2\vec{e}_3 = -\vec{e}_3\vec{e}_2 = \vec{g}_2 = q_{20}\vec{e}_0 + q_{21}\vec{e}_1 + q_{22}\vec{e}_2 + q_{23}\vec{e}_3 \\
&\vec{e}_3\vec{e}_1 = -\vec{e}_1\vec{e}_3 = \vec{g}_3 = q_{30}\vec{e}_0 + q_{31}\vec{e}_1 + q_{32}\vec{e}_2 + q_{33}\vec{e}_3
\end{aligned}
\tag{1.2}
$$

should be carried out.

Factors $p_{ij}$ and $q_{ij}$ are the subjects to be defined. We shall in addition impose on designed algebra a condition of associativity of multiplication:

$$(\vec{e}_i\vec{e}_j)\vec{e}_k = \vec{e}_i(\vec{e}_j\vec{e}_k) \quad \forall\, i, j, k = 0,1,2,3. \tag{1.3}$$

Multiplying from the left by $\vec{e}_0$ the first of chains of equalities (1.2) and using conditions (1.3), we shall obtain:

$$\vec{e}_0\vec{f}_0 = \vec{f}_0\vec{e}_0 = \vec{f}_1\vec{e}_1 = \vec{f}_2\vec{e}_2 = \vec{f}_3\vec{e}_3 \ .$$

Similarly, multiplying from the right by $\vec{e}_0$ the first of chains of equalities (1.2) and again using conditions (1.3), we shall obtain:

$$\vec{f}_0\vec{e}_0 = \vec{e}_0\vec{f}_0 = \vec{e}_1\vec{f}_1 = \vec{e}_2\vec{f}_2 = \vec{e}_3\vec{f}_3 \ .$$

Comparing two last chains of equalities, we can obtain a uniform chain:

$$\vec{e}_0\vec{f}_0 = \vec{e}_1\vec{f}_1 = \vec{e}_2\vec{f}_2 = \vec{e}_3\vec{f}_3 = \vec{f}_0\vec{e}_0 = \vec{f}_1\vec{e}_1 = \vec{f}_2\vec{e}_2 = \vec{f}_3\vec{e}_3 \ .$$

Further we shall operate in a similar way. I.e. we multiply from the left and from the right by $\vec{e}_i$ consecutively all chains of equalities (1.2). In the obtained sequence of chains of equalities the identical elements are found. Equating to each other those chains in which these elements meet, we shall obtain:

$$
\begin{array}{llllllll}
\vec{e}_0\vec{f}_0 = & \vec{e}_1\vec{f}_1 = & \vec{e}_2\vec{f}_2 = & \vec{e}_3\vec{f}_3 & = \vec{f}_0\vec{e}_0 = & \vec{f}_1\vec{e}_1 & = \vec{f}_2\vec{e}_2 & = \vec{f}_3\vec{e}_3 \\
\vec{e}_0\vec{f}_1 = & \vec{e}_1\vec{f}_0 = & -\vec{e}_2\vec{g}_1 = & \vec{e}_3\vec{g}_3 & = \vec{f}_1\vec{e}_0 = & \vec{f}_0\vec{e}_1 & = \vec{g}_1\vec{e}_2 & = -\vec{g}_3\vec{e}_3 \\
\vec{e}_0\vec{f}_2 = & \vec{e}_1\vec{g}_1 = & \vec{e}_2\vec{f}_0 = & -\vec{e}_3\vec{g}_2 & = \vec{f}_2\vec{e}_0 = & -\vec{g}_1\vec{e}_1 = & \vec{f}_0\vec{e}_2 & = \vec{g}_2\vec{e}_3 \\
\vec{e}_0\vec{f}_3 = & -\vec{e}_1\vec{g}_3 = & \vec{e}_2\vec{g}_2 = & \vec{e}_3\vec{f}_0 & = \vec{f}_3\vec{e}_0 = & \vec{g}_3\vec{e}_1 & = -\vec{g}_2\vec{e}_2 = & \vec{f}_0\vec{e}_3 \\
\vec{e}_0\vec{g}_1 = & \vec{e}_1\vec{f}_2 = & -\vec{e}_2\vec{f}_1 = & & = \vec{g}_1\vec{e}_0 = & -\vec{f}_2\vec{e}_1 = & \vec{f}_1\vec{e}_2 & \\
\vec{e}_0\vec{g}_2 = & = & \vec{e}_2\vec{f}_3 = & -\vec{e}_3\vec{f}_2 & = \vec{g}_2\vec{e}_0 = & & = -\vec{f}_3\vec{e}_2 = & \vec{f}_2\vec{e}_3 \\
\vec{e}_0\vec{g}_3 = & -\vec{e}_1\vec{f}_3 = & = & \vec{e}_3\vec{f}_1 & = \vec{g}_3\vec{e}_0 = & \vec{f}_3\vec{e}_1 = & & = -\vec{f}_1\vec{e}_3 \\
& \vec{e}_1\vec{g}_2 = & \vec{e}_2\vec{g}_3 = & \vec{e}_3\vec{g}_1 & = & = \vec{g}_2\vec{e}_1 & = \vec{g}_3\vec{e}_2 & = \vec{g}_1\vec{e}_3
\end{array}
\tag{1.4}
$$

Let's substitute in expression $\vec{e}_0\vec{f}_0$ the value of $\vec{f}_0$ obtained from the right part of first chain of equalities (1.2), and we shall make multiplication in all obtained expressions $\vec{e}_i\vec{e}_j$ . Then we shall obtain

$$\vec{e}_0\vec{f}_0 = \vec{e}_0(p_{00}\vec{e}_0 + p_{01}\vec{e}_1 + p_{02}\vec{e}_2 + p_{03}\vec{e}_3) =$$
$$= p_{00}\vec{f}_0 + p_{01}\vec{f}_1 + p_{02}\vec{f}_2 + p_{03}\vec{f}_3 .$$

Similarly we shall transform expression $\vec{f}_0\vec{e}_0$:

$$\vec{f}_0\vec{e}_0 = (p_{00}\vec{e}_0 + p_{01}\vec{e}_1 + p_{02}\vec{e}_2 + p_{03}\vec{e}_3)\vec{e}_0 =$$
$$= p_{00}\vec{f}_0 + p_{01}\vec{f}_1 + p_{02}\vec{f}_2 + p_{03}\vec{f}_3 .$$

From here it is visible, that equality $\vec{e}_0\vec{f}_0 = \vec{f}_0\vec{e}_0$ from the first chain of equalities (1.4) is satisfied identically.

In precisely the same way we investigate equality $\vec{e}_1\vec{f}_1 = \vec{f}_1\vec{e}_1$ from the same chain and we shall find out, what it will give us.

$$\vec{e}_1\vec{f}_1 = \vec{e}_1(p_{10}\vec{e}_0 + p_{11}\vec{e}_1 + p_{12}\vec{e}_2 + p_{13}\vec{e}_3) =$$
$$= p_{10}\vec{f}_1 + p_{11}\vec{f}_0 + p_{12}\vec{g}_1 - p_{13}\vec{g}_3 =$$
$$= \vec{f}_1\vec{e}_1 = (p_{10}\vec{e}_0 + p_{11}\vec{e}_1 + p_{12}\vec{e}_2 + p_{13}\vec{e}_3)\vec{e}_1 =$$
$$= p_{10}\vec{f}_1 + p_{11}\vec{f}_0 - p_{12}\vec{g}_1 + p_{13}\vec{g}_3$$

Comparing the second and the fourth lines in last expression, we find, that condition

$$p_{12}\vec{g}_1 - p_{13}\vec{g}_3 = 0$$

should be satisfied.

In a similar way having investigated all equalities of type $\vec{e}_i\vec{f}_j = \vec{f}_j\vec{e}_i$ or $\vec{e}_i\vec{g}_j = \vec{g}_j\vec{e}_i$ from (1.4), we shall obtain the following set of the conditions imposed on factors $p_{ij}$ and $q_{ij}$:

$$q_{10}\vec{f}_2 + q_{12}\vec{f}_0 = 0 \qquad p_{12}\vec{g}_1 - p_{13}\vec{g}_3 = 0$$
$$q_{30}\vec{f}_3 + q_{33}\vec{f}_0 = 0 \qquad p_{21}\vec{g}_1 - p_{23}\vec{g}_2 = 0$$
$$q_{10}\vec{f}_1 + q_{11}\vec{f}_0 = 0 \qquad p_{31}\vec{g}_3 - p_{32}\vec{g}_2 = 0$$
$$q_{20}\vec{f}_3 + q_{23}\vec{f}_0 = 0 \qquad p_{02}\vec{g}_1 - p_{03}\vec{g}_3 = 0$$
$$q_{30}\vec{f}_1 + q_{31}\vec{f}_0 = 0 \qquad p_{01}\vec{g}_1 - p_{03}\vec{g}_2 = 0$$
$$q_{20}\vec{f}_2 + q_{22}\vec{f}_0 = 0 \qquad p_{01}\vec{g}_3 - p_{02}\vec{g}_2 = 0$$
$$p_{20}\vec{f}_1 + p_{21}\vec{f}_0 = 0 \qquad q_{22}\vec{g}_1 - q_{23}\vec{g}_3 = 0$$
$$p_{10}\vec{f}_2 + p_{12}\vec{f}_0 = 0 \qquad q_{31}\vec{g}_1 - q_{33}\vec{g}_2 = 0$$
$$p_{30}\vec{f}_2 + p_{32}\vec{f}_0 = 0 \qquad q_{11}\vec{g}_3 - q_{12}\vec{g}_2 = 0$$

$$p_{20}\vec{f}_3 + p_{23}\vec{f}_0 = 0$$
$$p_{30}\vec{f}_1 + p_{31}\vec{f}_0 = 0$$
$$p_{10}\vec{f}_3 + p_{13}\vec{f}_0 = 0$$

From here it is easy to obtain a following set of chains of equalities:

$$\begin{aligned}
\vec{f}_1 &= (-\frac{q_{11}}{q_{10}})\vec{f}_0 = (-\frac{q_{31}}{q_{30}})\vec{f}_0 = (-\frac{p_{21}}{p_{20}})\vec{f}_0 = (-\frac{p_{31}}{p_{30}})\vec{f}_0 = -\alpha^1\vec{f}_0 \\
\vec{f}_2 &= (-\frac{q_{12}}{q_{10}})\vec{f}_0 = (-\frac{q_{22}}{q_{20}})\vec{f}_0 = (-\frac{p_{12}}{p_{10}})\vec{f}_0 = (-\frac{p_{32}}{p_{30}})\vec{f}_0 = -\alpha^2\vec{f}_0 \\
\vec{f}_3 &= (-\frac{q_{33}}{q_{30}})\vec{f}_0 = (-\frac{q_{23}}{q_{20}})\vec{f}_0 = (-\frac{p_{23}}{p_{20}})\vec{f}_0 = (-\frac{p_{13}}{p_{10}})\vec{f}_0 = -\alpha^3\vec{f}_0 \\
\vec{g}_3 &= (\frac{p_{12}}{p_{13}})\vec{g}_1 = (\frac{p_{02}}{p_{03}})\vec{g}_1 = (\frac{q_{22}}{q_{23}})\vec{g}_1 = \beta^3\vec{g}_1 \\
\vec{g}_2 &= (\frac{p_{21}}{p_{23}})\vec{g}_1 = (\frac{p_{01}}{p_{03}})\vec{g}_1 = (\frac{q_{31}}{q_{33}})\vec{g}_1 = \beta^2\vec{g}_1 \\
\vec{g}_3 &= (\frac{p_{32}}{p_{31}})\vec{g}_2 = (\frac{p_{02}}{p_{01}})\vec{g}_2 = (\frac{q_{12}}{q_{11}})\vec{g}_2 = \frac{\beta^3}{\beta^2}\vec{g}_1
\end{aligned} \quad . \qquad (1.5)$$

Comparing two expressions $\vec{f}_0 = p_{00}\vec{e}_0 + p_{01}\vec{e}_1 + p_{02}\vec{e}_2 + p_{03}\vec{e}_3$ and $\vec{f}_1 = p_{10}\vec{e}_0 + p_{11}\vec{e}_1 + p_{12}\vec{e}_2 + p_{13}\vec{e}_3$ and taking into account the first of equalities (1.5) $\vec{f}_1 = -\alpha^1\vec{f}_0$, we obtain:

$$\begin{aligned}
p_{10} &= -\alpha^1 p_{00} \\
p_{11} &= -\alpha^1 p_{01} \\
p_{12} &= -\alpha^1 p_{02} \\
p_{13} &= -\alpha^1 p_{03}
\end{aligned} \quad . \qquad (1.6)$$

Similarly it is possible to obtain

$$\begin{aligned}
p_{20} &= -\alpha^2 p_{00} & p_{30} &= -\alpha^3 p_{00} \\
p_{21} &= -\alpha^2 p_{01} & p_{31} &= -\alpha^3 p_{01} \\
p_{22} &= -\alpha^2 p_{02} & p_{32} &= -\alpha^3 p_{02} \\
p_{23} &= -\alpha^2 p_{03} & p_{33} &= -\alpha^3 p_{03} \\
\\
q_{20} &= \beta^2 q_{10} & q_{30} &= \beta^3 q_{10} \\
q_{21} &= \beta^2 q_{11} & q_{31} &= \beta^3 q_{11} \\
q_{22} &= \beta^2 q_{12} & q_{32} &= \beta^3 q_{12} \\
q_{23} &= \beta^2 q_{13} & q_{33} &= \beta^3 q_{13}
\end{aligned} \quad . \qquad (1.7)$$

After substitution of values of $p_{ij}$ and $q_{ij}$, taken from the right parts of the equalities (1.6-1.7) , into the chains of equalities (1.5) the following chains of equalities for factors $p_{ij}$ and $q_{ij}$ can be obtained:

$$-\frac{q_{11}}{q_{10}} = -\frac{\beta^3 q_{11}}{\beta^3 q_{10}} = -\frac{-\alpha^2 p_{01}}{-\alpha^2 p_{00}} = -\frac{-\alpha^3 p_{01}}{-\alpha^3 p_{00}} = -\alpha^1$$
$$-\frac{q_{12}}{q_{10}} = -\frac{\beta^2 q_{12}}{\beta^2 q_{10}} = -\frac{-\alpha^1 p_{02}}{-\alpha^1 p_{00}} = -\frac{-\alpha^3 p_{02}}{-\alpha^3 p_{00}} = -\alpha^2$$
$$-\frac{\beta^3 q_{13}}{\beta^3 q_{10}} = -\frac{\beta^2 q_{13}}{\beta^2 q_{10}} = -\frac{-\alpha^2 p_{03}}{-\alpha^2 p_{00}} = -\frac{-\alpha^1 p_{03}}{-\alpha^1 p_{00}} = -\alpha^3$$
$$-\frac{-\alpha^1 p_{02}}{-\alpha^1 p_{03}} = \frac{p_{02}}{p_{03}} = \frac{\beta^2 q_{12}}{\beta^2 q_{13}} = \beta^3$$
$$-\frac{-\alpha^2 p_{01}}{-\alpha^2 p_{03}} = \frac{p_{01}}{p_{03}} = \frac{\beta^3 q_{11}}{\beta^3 q_{13}} = \beta^2$$
$$-\frac{-\alpha^3 p_{02}}{-\alpha^3 p_{01}} = \frac{p_{02}}{p_{01}} = \frac{q_{12}}{q_{11}} = \frac{\beta^3}{\beta^2}$$

or

$$\frac{q_{11}}{q_{10}} = \frac{p_{01}}{p_{00}} = \alpha^1$$
$$\frac{q_{12}}{q_{10}} = \frac{p_{02}}{p_{00}} = \alpha^2$$
$$\frac{q_{13}}{q_{10}} = \frac{p_{03}}{p_{00}} = \alpha^3 \quad . \qquad (1.8)$$
$$\frac{p_{02}}{p_{03}} = \frac{q_{12}}{q_{13}} = \beta^3$$
$$\frac{p_{01}}{p_{03}} = \frac{q_{11}}{q_{13}} = \beta^2$$

From here parities

$$p_{01} = \alpha^1 p_{00} \qquad q_{11} = \alpha^1 q_{10}$$
$$p_{02} = \alpha^2 p_{00} \qquad q_{12} = \alpha^2 q_{10} \quad . \qquad (1.9)$$
$$p_{03} = \alpha^3 p_{00} \qquad q_{13} = \alpha^3 q_{10}$$

turn out.

Taking into account these parities, and the last two of equalities (1.8) as well, we obtain

$$\frac{p_{02}}{p_{03}} = \frac{\alpha^2 p_{00}}{\alpha^3 p_{00}} = \frac{\alpha^2}{\alpha^3} = \beta^3$$
$$\frac{p_{01}}{p_{03}} = \frac{\alpha^1 p_{00}}{\alpha^3 p_{00}} = \frac{\alpha^1}{\alpha^3} = \beta^2 \quad .$$

Further, taking into account last parities and parities (1.5) and (1.9), we obtain following expressions for vectors $\vec{f}$ and $\vec{g}_i$:

$$\vec{f}_0 = p_{00}(e_0 + \alpha^1 e_1 + \alpha^2 e_2 + \alpha^3 e_3)$$
$$\vec{f}_1 = -\alpha^1 p_{00}(e_0 + \alpha^1 e_1 + \alpha^2 e_2 + \alpha^3 e_3)$$
$$\vec{f}_2 = -\alpha^2 p_{00}(e_0 + \alpha^1 e_1 + \alpha^2 e_2 + \alpha^3 e_3)$$
$$\vec{f}_3 = -\alpha^3 p_{00}(e_0 + \alpha^1 e_1 + \alpha^2 e_2 + \alpha^3 e_3)$$
$$\vec{g}_1 = q_{10}(e_0 + \alpha^1 e_1 + \alpha^2 e_2 + \alpha^3 e_3)$$
$$\vec{g}_2 = \frac{\alpha^1}{\alpha^3} q_{10}(e_0 + \alpha^1 e_1 + \alpha^2 e_2 + \alpha^3 e_3)$$
$$\vec{g}_3 = \frac{\alpha^2}{\alpha^3} q_{10}(e_0 + \alpha^1 e_1 + \alpha^2 e_2 + \alpha^3 e_3)$$

If we shall enter following designations:

$$\begin{aligned} p_{00} &= a \\ q_{10} &= \alpha^3 c \\ \vec{e}_0 + \alpha^1 \vec{e}_1 + \alpha^2 \vec{e}_2 + \alpha^3 \vec{e}_3 &= \vec{R} \end{aligned} \quad , \qquad (1.10)$$

then expressions for vectors $\vec{f}_i$ and $\vec{g}_i$ get a kind:

$$\begin{aligned} \vec{f}_0 &= \vec{e}_0\vec{e}_0 = \vec{e}_1\vec{e}_1 = \vec{e}_2\vec{e}_2 = \vec{e}_3\vec{e}_3 = a\vec{R} \\ \vec{f}_1 &= \vec{e}_1\vec{e}_0 = \vec{e}_0\vec{e}_1 = -\alpha^1 a\vec{R} \\ \vec{f}_2 &= \vec{e}_2\vec{e}_0 = \vec{e}_0\vec{e}_2 = -\alpha^2 a\vec{R} \\ \vec{f}_3 &= \vec{e}_3\vec{e}_0 = \vec{e}_0\vec{e}_3 = -\alpha^3 a\vec{R} \\ \vec{g}_1 &= \vec{e}_1\vec{e}_2 = -\vec{e}_2\vec{e}_1 = \alpha^3 c\vec{R} \\ \vec{g}_2 &= \vec{e}_2\vec{e}_3 = -\vec{e}_3\vec{e}_2 = \alpha^1 c\vec{R} \\ \vec{g}_3 &= \vec{e}_3\vec{e}_1 = -\vec{e}_1\vec{e}_3 = \alpha^2 c\vec{R} \end{aligned} \quad . \qquad (1.11)$$

From here it is easy to obtain parities:

$$\begin{aligned} \vec{e}_0\vec{R} &= (1-\alpha^{1^2}-\alpha^{2^2}-\alpha^{3^2})a\vec{R} & \vec{R}\vec{e}_0 &= (1-\alpha^{1^2}-\alpha^{2^2}-\alpha^{3^2})a\vec{R} \\ \vec{e}_1\vec{R} &= 0 & \vec{R}\vec{e}_1 &= 0 \\ \vec{e}_2\vec{R} &= 0 & \vec{R}\vec{e}_2 &= 0 \\ \vec{e}_3\vec{R} &= 0 & \vec{R}\vec{e}_3 &= 0 \end{aligned}$$

Further from here it is possible to obtain

$\vec{e}_0 \vec{f}_0 = (1 - \alpha^{1^2} - \alpha^{2^2} - \alpha^{3^2}) a\vec{R}$ and $\vec{e}_1 \vec{f}_1 = 0$.

But from parities (1.4) it is known, that $\vec{e}_0 \vec{f}_0 = \vec{e}_1 \vec{f}_1$, whence at once follows the condition

$$1\text{-}\alpha^{1^2}\text{-}\alpha^{2^2}\text{-}\alpha^{3^2} = 0 \ . \qquad (1.12)$$

It is obvious, that for any $i, j, k = 0,...,3$ the condition

$$(\vec{e}_i \vec{e}_j)\vec{e}_k = \vec{e}_i(\vec{e}_j \vec{e}_k) = 0$$

is satisfied.
It is clear, that any (in the sense of arrangement of brackets) product of any three elements in this algebra is equal to zero.
So, the following **Theorem 1.1** is proved:
The associative algebra in four-dimensional linear space with the special law (1.1) of multiplication of any element by its conjugated possesses the following properties: products of basic elements $\vec{e}_i$ submit to parities (1.11) and are proportional to the same element $\vec{R}$ set by third of parities (1.10). In parities (1.11) $a$ and $c$ are arbitrary real constants, and constants $\alpha^i$ are real and submit to conditions (1.12). Obviously, the product of any two elements is proportional to $\vec{R}$, and any product in the sense of arrangement of brackets of any three elements to equal to zero.
**The remark**. It is obvious, that the designed algebra cannot be named algebra of hypercomplex numbers as by virtue of parities (1.11) no one its subalgebra is not an algebra of real numbers.
We shall investigate the further properties of the obtained algebra.
Let $\vec{r} = x^0\vec{e}_0 + x^1\vec{e}_1 + x^2\vec{e}_2 + x^3\vec{e}_3$ be an arbitrary element. Does such uniquely certain element $\vec{E} = y^0\vec{e}_0 + y^1\vec{e}_1 + y^2\vec{e}_2 + y^3\vec{e}_3$ that the parity $\vec{r}\vec{E} = \vec{r}$ is carried out exist?

$$\begin{aligned}\vec{r}\vec{E} &= (x^0\vec{e}_0 + x^1\vec{e}_1 + x^2\vec{e}_2 + x^3\vec{e}_3)(y^0\vec{e}_0 + y^1\vec{e}_1 + y^2\vec{e}_2 + y^3\vec{e}_3) = \\ &= [(x^0y^0 + x^1y^1 + x^2y^2 + x^3y^3)a + (-x^1y^0 - x^0y^1)\alpha^1 a + \\ &+ (-x^2y^0 - x^0y^2)\alpha^2 a + (-x^3y^0 - x^0y^3)\alpha^3 a + (-x^2y^1 + x^1y^2)\alpha^3 c + \\ &+ (x^3y^1 - x^1y^3)\alpha^2 c + (-x^3y^2 + x^2y^3)\alpha^1 a](\vec{e}_0 + \alpha^1\vec{e}_1 + \alpha^2\vec{e}_2 + \alpha^3\vec{e}_3)\end{aligned}$$

But this expression should be equal to $\vec{r} = x^0\vec{e}_0 + x^1\vec{e}_1 + x^2\vec{e}_2 + x^3\vec{e}_3$. If to designate by $A$ the expression in square brackets from the last equality the parities

$$\begin{aligned} A &= x^0 \\ A\alpha^1 &= x^1 \\ A\alpha^2 &= x^2 \\ A\alpha^3 &= x^3 \end{aligned}$$

should be carried out.
From here by virtue of randomness of coordinates $x^i$ and randomness, but fixity for the given algebra of constants $\alpha^j$, the impossibility of performance of last four equalities follows. So, the **Theorem 1.2** is proved: the investigated algebra is algebra without unity.

We shall name right quotient from division of an element $\vec{r} = x^0\vec{e}_0 + x^1\vec{e}_1 + x^2\vec{e}_2 + x^3\vec{e}_3$ by element $\vec{\rho} = y^0\vec{e}_0 + y^1\vec{e}_1 + y^2\vec{e}_2 + y^3\vec{e}_3$ the element $\vec{\eta} = z^0\vec{e}_0 + z^1\vec{e}_1 + z^2\vec{e}_2 + z^3\vec{e}_3$: $\{\frac{\vec{r}}{\vec{\rho}}\}_r = \vec{\eta}$, if $\vec{r} = \vec{\eta}\vec{\rho}$. Having made multiplication in the last equality, we shall obtain:

$$\begin{aligned}
\vec{r} &= x^0\vec{e}_0 + x^1\vec{e}_1 + x^2\vec{e}_2 + x^3\vec{e}_3 = \\
&= \vec{\eta}\vec{\rho} = (z^0\vec{e}_0 + z^1\vec{e}_1 + z^2\vec{e}_2 + z^3\vec{e}_3)(y^0\vec{e}_0 + y^1\vec{e}_1 + y^2\vec{e}_2 + y^3\vec{e}_3) = \\
&= [(z^0y^0 + z^1y^1 + z^2y^2 + z^3y^3)a + (-z^1y^0 - z^0y^1)\alpha^1 a + \\
&+ (-z^2y^0 - z^0y^2)\alpha^2 a + (-z^3y^0 - z^0y^3)\alpha^3 a + (-z^2y^1 + z^1y^2)\alpha^3 c + \\
&+ (z^3y^1 - z^1y^3)\alpha^2 c + (-z^3y^2 + z^2y^3)\alpha^1 a](\vec{e}_0 + \alpha^1\vec{e}_1 + \alpha^2\vec{e}_2 + \alpha^3\vec{e}_3)
\end{aligned}.$$

If to designate expression in square brackets as $A$ parities

$$\begin{aligned}
A &= x^0 \\
A\alpha^1 &= x^1 \\
A\alpha^2 &= x^2 \\
A\alpha^3 &= x^3
\end{aligned} \tag{1.13}$$

should be carried out.
From here it is visible that, generally speaking, it is impossible to divide from the right arbitrary element by another arbitrary. It is obvious, that it is possible to divide only the elements proportional to $\vec{R}$, and they can be divided by arbitrary elements, but $\vec{R}$. In this case the last four equalities are carried out. But it is impossible to define right quotient from division unequivocally, since from one equation $A = x^0$ (other three equations are proportional to the first) it is impossible to define unequivocally four unknown quantities $z^i$.
So, the **Teorem 1.3** is proved:
the investigated algebra is algebra without division.
**The remark**. It is easy to notice, that if $\vec{\rho} = \vec{R}$, then $A = 0$, i.e. the zero element is the only element that can be divided by element proportional to $\vec{R}$ from the right. And in general all elements of this algebra are divisors of zero.
All aforesaid appears absolutely similar if to define left quotient from division of elements: $\{\frac{\vec{r}}{\vec{\rho}}\}_l = \vec{\eta}$, if $\vec{r} = \vec{\rho}\vec{\eta}$.

## 2. The Analysis in Flat Space.

Now we shall try to find out, is it possible to construct any acceptable analysis using algebra with such properties. Let $\vec{u} = u^0\vec{e}_0 + u^1\vec{e}_1 + u^2\vec{e}_2 + u^3\vec{e}_3$ be a four-dimensional function of four-dimensional argument $\vec{r} = x^0\vec{e}_0 + x^1\vec{e}_1 + x^2\vec{e}_2 + x^3\vec{e}_3$. We shall name the left derivative of function $\vec{u}$ of argument $\vec{r}$ in a point $\vec{r}$ a limit (if it, certainly, exists) of difference expression $\{\frac{\vec{u}(\vec{r} + \Delta\vec{r}) - \vec{u}(\vec{r})}{\Delta\vec{r}}\}_l$. This limit we shall designate $\vec{u}'_l(\vec{r})$, i.e.

$$\vec{u}'_l(\vec{r}) = \lim_{\Delta\vec{r}\to 0}\{\frac{\vec{u}(\vec{r} + \Delta\vec{r}) - \vec{u}(\vec{r})}{\Delta\vec{r}}\}_l. \tag{2.1}$$

As a limit of sequence of vectors $\vec{r}_n = x_n^0\vec{e}_0 + x_n^1\vec{e}_1 + x_n^2\vec{e}_2 + x_n^3\vec{e}_3$ we shall understand a vector $\vec{r} = x^0\vec{e}_0 + x^1\vec{e}_1 + x^2\vec{e}_2 + x^3\vec{e}_3 = (\lim_{n\to\infty} x_n^0)\vec{e}_0 + (\lim_{n\to\infty} x_n^1)\vec{e}_1 + (\lim_{n\to\infty} x_n^2)\vec{e}_2 + (\lim_{n\to\infty} x_n^3)\vec{e}_3$. The limit, certainly, shall not depend on a way for $\Delta\vec{r}$ to reach zero. Let's enter designations $\vec{u}_l'(\vec{r}) = v^0\vec{e}_0 + v^1\vec{e}_1 + v^2\vec{e}_2 + v^3\vec{e}_3$. Now we shall derive an analogue of conditions of Cauchy-Riemann. Let be $\Delta\vec{r} = \Delta x^0\vec{e}_0$.
Then

$$\begin{aligned}
\vec{u}_l'(\vec{r}) = \lim_{\Delta x^0\to\infty} [&\frac{u^0(x^0+\Delta x^0, x^1, x^2, x^3) - u^0(x^0, x^1, x^2, x^3)}{\Delta x^0}\vec{e}_0 + \\
&+\frac{u^1(x^0+\Delta x^0, x^1, x^2, x^3) - u^1(x^0, x^1, x^2, x^3)}{\Delta x^0}\vec{e}_1 + \\
&+\frac{u^2(x^0+\Delta x^0, x^1, x^2, x^3) - u^2(x^0, x^1, x^2, x^3)}{\Delta x^0}\vec{e}_2 + \\
&+\frac{u^3(x^0+\Delta x^0, x^1, x^2, x^3) - u^3(x^0, x^1, x^2, x^3)}{\Delta x^0}\vec{e}_3 ] / \vec{e}_0 = \\
= &\frac{u_{x^0}^0(x^0,x^1,x^2,x^3)\vec{e}_0 + u_{x^0}^1(x^0,x^1,x^2,x^3)\vec{e}_1 + u_{x^0}^2(x^0,x^1,x^2,x^3)\vec{e}_2 + u_{x^0}^3(x^0,x^1,x^2,x^3)\vec{e}_3}{\vec{e}_0} = \\
= &v^0(x^0,x^1,x^2,x^3)\vec{e}_0 + v^1(x^0,x^1,x^2,x^3)\vec{e}_1 + v^2(x^0,x^1,x^2,x^3)\vec{e}_2 + v^3(x^0,x^1,x^2,x^3)\vec{e}_3
\end{aligned}$$

Here $u_{x^i}^0(x^0,x^1,x^2,x^3)$ means a derivative of function $u^0(x^0,x^1,x^2,x^3)$ along argument $x^i$. Everywhere further, where it will not cause doubts, we shall omit the designations of arguments of functions.
The numerator of the expression in a penultimate line, by virtue of the theorem 1.3 is obliged to look like (since a vector in a denominator $\vec{e}_0 \neq \vec{R}$):

$$u_{x^0}^0\vec{e}_0 + u_{x^0}^1\vec{e}_1 + u_{x^0}^2\vec{e}_2 + u_{x^0}^3\vec{e}_3 = A\vec{R} = A(\vec{e}_0 + \alpha^1\vec{e}_1 + \alpha^2\vec{e}_2 + \alpha^3\vec{e}_3) \quad \text{or}$$

$$\begin{aligned}
u_{x^0}^0 &= A \\
u_{x^0}^1 &= \alpha^1 A \\
u_{x^0}^2 &= \alpha^2 A \\
u_{x^0}^3 &= \alpha^3 A
\end{aligned}.$$

From here we obtain:

$$\begin{aligned}
u_{x^0}^1 &= \alpha^1 u_{x^0}^0 \\
u_{x^0}^2 &= \alpha^2 u_{x^0}^0 \\
u_{x^0}^3 &= \alpha^3 u_{x^0}^0
\end{aligned}.$$

On the other hand, having taken into account the definition of left quotient from division of one element by another and the rules of multiplication of basic vectors, we shall obtain:

$$A\vec{R} = \vec{e}_0(v^0\vec{e}_0 + v^1\vec{e}_1 + v^2\vec{e}_2 + v^3\vec{e}_3) = (v^0 - \alpha^1 v^1 - \alpha^2 v^2 - \alpha^3 v^3)a\vec{R} \ ,$$

whence we obtain one of the equations connecting components of derivatives with derivatives of components of function:

$$u^0_{x^0} = (v^0 - \alpha^1 v^1 - \alpha^2 v^2 - \alpha^3 v^3)a \ .$$

Similarly, having taken as a vector $\Delta\vec{r}$ vectors $\Delta x^1\vec{e}_1$, $\Delta x^2\vec{e}_2$ and $\Delta x^3\vec{e}_3$, we shall obtain the following analogues of conditions of Cauchy-Riemann:

$$\begin{array}{lll} u^1_{x^1} = \alpha^1 u^0_{x^1} & u^1_{x^2} = \alpha^1 u^0_{x^2} & u^1_{x^3} = \alpha^1 u^0_{x^3} \\ u^2_{x^1} = \alpha^2 u^0_{x^1} & u^2_{x^2} = \alpha^2 u^0_{x^2} & u^2_{x^3} = \alpha^2 u^0_{x^3} \ , \\ u^3_{x^1} = \alpha^3 u^0_{x^1} & u^3_{x^2} = \alpha^3 u^0_{x^2} & u^3_{x^3} = \alpha^3 u^0_{x^3} \end{array}$$

and the further equations connecting components of derivatives with derivatives of components of function

$$\begin{array}{l} u^0_{x^1} = -\alpha^1 a v^0 + a v^1 + \alpha^3 c v^2 - \alpha^2 c v^3 \\ u^0_{x^2} = -\alpha^2 a v^0 - \alpha^3 c v^1 + a v^2 + \alpha^1 c v^3 \ . \\ u^0_{x^3} = -\alpha^3 a v^0 + \alpha^2 c v^1 - \alpha^1 c v^2 + a v^3 \end{array}$$

At last, let

$$\begin{array}{l} \Delta\vec{r} = B\vec{R} = \Delta x^0\vec{e}_0 + \Delta x^1\vec{e}_1 + \Delta x^2\vec{e}_2 + \Delta x^3\vec{e}_3 = \\ \quad = \Delta x^0\left(\vec{e}_0 + \dfrac{\Delta x^1}{\Delta x^0}\vec{e}_1 + \dfrac{\Delta x^2}{\Delta x^0}\vec{e}_2 + \dfrac{\Delta x^3}{\Delta x^0}\vec{e}_3\right) \end{array} \ ,$$

whence

$$\begin{array}{l} \Delta x^0 = B \\ \dfrac{\Delta x^1}{\Delta x^0} = \alpha^1 \\ \dfrac{\Delta x^2}{\Delta x^0} = \alpha^2 \\ \dfrac{\Delta x^3}{\Delta x^0} = \alpha^3 \end{array}$$

and

$$\begin{array}{l} \Delta x^1 = \alpha^1 \Delta x^0 \\ \Delta x^2 = \alpha^2 \Delta x^0 \\ \Delta x^3 = \alpha^3 \Delta x^0 \end{array}$$

can be obtained.

And for the value of the derivative the following expression is obtained:

$$\vec{u}_l'(\vec{r}) = \lim_{\Delta x^0 \to \infty} [\frac{u^0(x^0+\Delta x^0, x^1+\alpha^1\Delta x^0, x^2+\alpha^2\Delta x^0, x^3+\alpha^3\Delta x^0) - u^0(x^0,x^1,x^2,x^3)}{\Delta x^0}\vec{e}_0 +$$
$$+\frac{u^1(x^0+\Delta x^0, x^1+\alpha^1\Delta x^0, x^2+\alpha^2\Delta x^0, x^3+\alpha^3\Delta x^0) - u^1(x^0,x^1,x^2,x^3)}{\Delta x^0}\vec{e}_1 +$$
$$+\frac{u^2(x^0+\Delta x^0, x^1+\alpha^1\Delta x^0, x^2+\alpha^2\Delta x^0, x^3+\alpha^3\Delta x^0) - u^2(x^0,x^1,x^2,x^3)}{\Delta x^0}\vec{e}_2 +$$
$$+\frac{u^3(x^0+\Delta x^0, x^1+\alpha^1\Delta x^0, x^2+\alpha^2\Delta x^0, x^3+\alpha^3\Delta x^0) - u^3(x^0,x^1,x^2,x^3)}{\Delta x^0}\vec{e}_3]/\vec{R} = .$$
$$= \{[u^0_{x^0}(x^0,x^1,x^2,x^3) + \alpha^1 u^0_{x^1}(x^0,x^1,x^2,x^3) + \alpha^2 u^0_{x^2}(x^0,x^1,x^2,x^3) + \alpha^3 u^0_{x^3}(x^0,x^1,x^2,x^3)]\vec{e}_0 +$$
$$+[u^1_{x^0}(x^0,x^1,x^2,x^3) + \alpha^1 u^1_{x^1}(x^0,x^1,x^2,x^3) + \alpha^2 u^1_{x^2}(x^0,x^1,x^2,x^3) + \alpha^3 u^1_{x^3}(x^0,x^1,x^2,x^3)]\vec{e}_1 +$$
$$+[u^2_{x^0}(x^0,x^1,x^2,x^3) + \alpha^1 u^2_{x^1}(x^0,x^1,x^2,x^3) + \alpha^2 u^2_{x^2}(x^0,x^1,x^2,x^3) + \alpha^3 u^2_{x^3}(x^0,x^1,x^2,x^3)]\vec{e}_2 +$$
$$+[u^3_{x^0}(x^0,x^1,x^2,x^3) + \alpha^1 u^3_{x^1}(x^0,x^1,x^2,x^3) + \alpha^2 u^3_{x^2}(x^0,x^1,x^2,x^3) + \alpha^3 u^3_{x^3}(x^0,x^1,x^2,x^3)]\vec{e}_3\}/\vec{R}$$

As the denominator of last expression is $\vec{R}$, according to the remark to the theorem 1.3 numerator is obliged to be a zero vector. From here, receiving the remained analogues of conditions of Cauchy-Riemann, we write down them together with already available:

$$\begin{aligned}
&u^0_{x^0} + \alpha^1 u^0_{x^1} + \alpha^2 u^0_{x^2} + \alpha^3 u^0_{x^3} = 0\\
&u^1_{x^0} + \alpha^1 u^1_{x^1} + \alpha^2 u^1_{x^2} + \alpha^3 u^1_{x^3} = 0\\
&u^2_{x^0} + \alpha^1 u^2_{x^1} + \alpha^2 u^2_{x^2} + \alpha^3 u^2_{x^3} = 0\\
&u^3_{x^0} + \alpha^1 u^3_{x^1} + \alpha^2 u^3_{x^2} + \alpha^3 u^3_{x^3} = 0\\
&u^1_{x^0} = \alpha^1 u^0_{x^0} \quad u^1_{x^1} = \alpha^1 u^0_{x^1} \quad u^1_{x^2} = \alpha^1 u^0_{x^2} \quad u^1_{x^3} = \alpha^1 u^0_{x^3}\\
&u^2_{x^0} = \alpha^2 u^0_{x^0} \quad u^2_{x^1} = \alpha^2 u^0_{x^1} \quad u^2_{x^2} = \alpha^2 u^0_{x^2} \quad u^2_{x^3} = \alpha^2 u^0_{x^3}\\
&u^3_{x^0} = \alpha^3 u^0_{x^0} \quad u^3_{x^1} = \alpha^3 u^0_{x^1} \quad u^3_{x^2} = \alpha^3 u^0_{x^2} \quad u^3_{x^3} = \alpha^3 u^0_{x^3}
\end{aligned} \tag{2.2}$$

Besides we shall write down together all the equations connecting components of derivatives with derivatives of components of function:

$$\begin{aligned}
&u^0_{x^0} = (v^0 - \alpha^1 v^1 - \alpha^2 v^2 - \alpha^3 v^3)a\\
&u^0_{x^1} = -\alpha^1 a v^0 + a v^1 + \alpha^3 c v^2 - \alpha^2 c v^3\\
&u^0_{x^2} = -\alpha^2 a v^0 - \alpha^3 c v^1 + a v^2 + \alpha^1 c v^3\\
&u^0_{x^3} = -\alpha^3 a v^0 + \alpha^2 c v^1 - \alpha^1 c v^2 + a v^3
\end{aligned} . \tag{2.3}$$

System of the equations (2.3) for components of a derivative has a lot of solutions. This fact is easy to be understood, if you multiply second of them by $\alpha^1$, third – by $\alpha^2$ and fourth – by $\alpha^3$, and add all four term by term. Thus the identity $0 = 0$ turns out. It means that it is impossible unequivocally to express components of a derivative through derivatives of components of a vector-function. However, it is possible to obtain expressions for the components of the left derivative $v^1$, $v^2$, $v^3$, having expressed them through $v^0$. The expression for a vector-function of the left derivative is:

$$\vec{u}'_l = v^0\vec{e}_0 + [\alpha^1 v^0 + \frac{a^2 u^0_{x^1} + ac(\alpha^2 u^0_{x^3} - \alpha^3 u^0_{x^2}) - c^2\alpha^1 u^0_{x^0}}{a(a^2+c^2)}]\vec{e}_1 + $$
$$+[\alpha^2 v^0 + \frac{a^2 u^0_{x^2} + ac(\alpha^3 u^0_{x^1} - \alpha^1 u^0_{x^3}) - c^2\alpha^2 u^0_{x^0}}{a(a^2+c^2)}]\vec{e}_2 + \quad , \qquad (2.4)$$
$$+[\alpha^3 v^0 + \frac{a^2 u^0_{x^3} + ac(\alpha^1 u^0_{x^2} - \alpha^2 u^0_{x^1}) - c^2\alpha^3 u^0_{x^0}}{a(a^2+c^2)}]\vec{e}_3$$

where $v^0$ is arbitrary function.
What about system (2.2), this system as system of the linear equations for private derivatives of components of a vector-function has a lot of solutions as well. The system of the independent equations can look, for example, so:

$$\begin{aligned}
&u^0_{x^0} = -u^0_{x^1}\alpha^1 - u^0_{x^2}\alpha^2 - u^0_{x^3}\alpha^3 \\
&u^1_{x^0} = -\alpha^1(u^0_{x^1}\alpha^1 + u^0_{x^2}\alpha^2 + u^0_{x^3}\alpha^3) \\
&u^2_{x^0} = -\alpha^2(u^0_{x^1}\alpha^1 + u^0_{x^2}\alpha^2 + u^0_{x^3}\alpha^3) \\
&u^3_{x^0} = -\alpha^3(u^0_{x^1}\alpha^1 + u^0_{x^2}\alpha^2 + u^0_{x^3}\alpha^3) \\
&u^1_{x^1} = \alpha^1 u^0_{x^1} \qquad u^1_{x^2} = \alpha^1 u^0_{x^2} \qquad u^1_{x^3} = \alpha^1 u^0_{x^3} \\
&u^2_{x^1} = \alpha^2 u^0_{x^1} \qquad u^2_{x^2} = \alpha^2 u^0_{x^2} \qquad u^2_{x^3} = \alpha^2 u^0_{x^3} \\
&u^3_{x^1} = \alpha^3 u^0_{x^1} \qquad u^3_{x^2} = \alpha^3 u^0_{x^2} \qquad u^3_{x^3} = \alpha^3 u^0_{x^3}
\end{aligned} \quad , \qquad (2.5)$$

where all other private derivatives are expressed through $u^0_{x^1}$, $u^0_{x^2}$, $u^0_{x^3}$ . The system (2.2) in comparison with system (2.5) has more symmetrical kind.
Simple calculations allow to solve system (2.5) and to concretize a kind of differentiable functions:

$$\begin{aligned}
&u^0 = \mathrm{A}^0(x^1 - \alpha^1 x^0, x^2 - \alpha^2 x^0, x^3 - \alpha^3 x^0) \\
&u^1 = \alpha^1 \mathrm{A}^0(x^1 - \alpha^1 x^0, x^2 - \alpha^2 x^0, x^3 - \alpha^3 x^0) \\
&u^2 = \alpha^2 \mathrm{A}^0(x^1 - \alpha^1 x^0, x^2 - \alpha^2 x^0, x^3 - \alpha^3 x^0) \\
&u^3 = \alpha^3 \mathrm{A}^0(x^1 - \alpha^1 x^0, x^2 - \alpha^2 x^0, x^3 - \alpha^3 x^0)
\end{aligned} \quad , \qquad (2.6)$$

where $\mathrm{A}^0$ is an arbitrary function of three specified arguments.
If now to repeat all course of reasonings for the right derivative it is possible to obtain the analogue of the formula (2.4) for it:

$$\vec{u}'_r = w^0\vec{e}_0 + [\alpha^1 w^0 + \frac{a^2 u^0_{x^1} - ac(\alpha^2 u^0_{x^3} - \alpha^3 u^0_{x^2}) - c^2\alpha^1 u^0_{x^0}}{a(a^2+c^2)}]\vec{e}_1 + $$
$$+[\alpha^2 w^0 + \frac{a^2 u^0_{x^2} - ac(\alpha^3 u^0_{x^1} - \alpha^1 u^0_{x^3}) - c^2\alpha^2 u^0_{x^0}}{a(a^2+c^2)}]\vec{e}_2 + \quad . \qquad (2.7)$$
$$+[\alpha^3 w^0 + \frac{a^2 u^0_{x^3} - ac(\alpha^1 u^0_{x^2} - \alpha^2 u^0_{x^1}) - c^2\alpha^3 u^0_{x^0}}{a(a^2+c^2)}]\vec{e}_3$$

Let's name function differentiable if its right and left derivatives exist and are equal. Then, comparing formulas (2.4) and (2.7), we obtain

$$w^0 = v^0 \ .$$

Besides we obtain the additional conditions imposed on a kind of function $\mathrm{A}^0$ :

$$\begin{aligned} \alpha^2 u^0_{x^3} - \alpha^3 u^0_{x^2} &= 0 \\ \alpha^3 u^0_{x^1} - \alpha^1 u^0_{x^3} &= 0 \\ \alpha^1 u^0_{x^2} - \alpha^2 u^0_{x^1} &= 0 \end{aligned} \ .$$

Having taken into account the first of formulas (2.6) and using last formulas, again by small calculations we concretize a kind of function even more. Then for the components of differentiable function the following expressions are obtained:

$$\begin{aligned} u^0 &= \mathrm{A}^0(x^0 - \alpha^1 x^1 - \alpha^2 x^2 - \alpha^3 x^3) \\ u^1 &= \alpha_1 \mathrm{A}^0(x^0 - \alpha^1 x^1 - \alpha^2 x^2 - \alpha^3 x^3) \\ u^2 &= \alpha_2 \mathrm{A}^0(x^0 - \alpha^1 x^1 - \alpha^2 x^2 - \alpha^3 x^3) \\ u^3 &= \alpha_3 \mathrm{A}^0(x^0 - \alpha^1 x^1 - \alpha^2 x^2 - \alpha^3 x^3) \end{aligned} \ . \qquad (2.8)$$

So, the **Theorem (2.1)** is proved:
Components of differentiable function satisfy the system of the differential equations (2.2) and look like (2.8), where $\mathrm{A}^0$ is an arbitrary function. Components of derivative of differentiable function are defined to within an arbitrary function $v^0$ and are given by the formula:

$$\begin{aligned} \vec{u}' &= v^0 \vec{e}_0 + [\alpha^1 v^0 + \frac{a^2 u^0_{x^1} - c^2 \alpha^1 u^0_{x^0}}{a(a^2 + c^2)}] \vec{e}_1 + \\ &\quad + [\alpha^2 v^0 + \frac{a^2 u^0_{x^2} - c^2 \alpha^2 u^0_{x^0}}{a(a^2 + c^2)}] \vec{e}_2 + \\ &\quad + [\alpha^3 v^0 + \frac{a^2 u^0_{x^3} - c^2 \alpha^3 u^0_{x^0}}{a(a^2 + c^2)}] \vec{e}_3 = \\ &= v^0 \vec{e}_0 + \alpha^1 [v^0 - \frac{A_0'}{a}] \vec{e}_1 + \alpha^2 [v^0 - \frac{A_0'}{a}] \vec{e}_2 + \alpha^3 [v^0 - \frac{A_0'}{a}] \vec{e}_3 \end{aligned} \ , \qquad (2.9)$$

where $A_0'$ means a derivative of function $A_0$ along all complex argument.

**The remark 1**. Function $\tilde{\tilde{u}} = \tilde{u}^0 \vec{e}_0 + \alpha^1 \tilde{u}^0 \vec{e}_1 + \alpha^2 \tilde{u}^0 \vec{e}_2 + \alpha^3 \tilde{u}^0 \vec{e}_3$ where $\tilde{u}^0 = \tilde{\mathrm{A}}^0(x^0 + \alpha^1 x^1 + \alpha^2 x^2 + \alpha^3 x^3)$ is not differentiable.

**The remark 2**. The derivative of function is not differentiable, because its components do not satisfy (2.8).

## 3. Linear Transformations of Basis and Coordinates.

We shall find out which linear transformation of basis and coordinates keeps a kind (1.1) for product of a vector and its conjugated $\vec{r}\vec{r}'$.

Let $(a_i^j)$ be a matrix of transformation of coordinates: $x^j = \sum_i y^i a_i^j$ (summation, certainly, is conducted from 0 up to 3); $(b_i^j)$ is a matrix of transformation of basis: $\vec{e}_i = \sum_i b_i^j \vec{k}_j$. We shall demand the equality

$$\vec{r}\vec{r}' = ((x^0)^2 - (x^1)^2 - (x^2)^2 - (x^3)^2)\vec{f}_0 = ((y^0)^2 - (y^1)^2 - (y^2)^2 - (y^3)^2)\vec{\varphi}_0$$

to be carried out.
Then

$$\begin{aligned}\vec{r}\vec{r}' &= ((x^0)^2 - (x^1)^2 - (x^2)^2 - (x^3)^2)\vec{f}_0 = \\ &= [(\sum_i y^i a_i^0)^2 - (\sum_i y^i a_i^1)^2 - (\sum_i y^i a_i^2)^2 - (\sum_i y^i a_i^3)^2]\vec{f}_0 = \\ &= [\sum_i y^i y^j (a_i^0 a_j^0 - a_i^1 a_j^1 - a_i^2 a_j^2 - a_i^3 a_j^3)]\vec{f}_0 = \\ &= (\sum_i y^i y^j c_{ij})\vec{f}_0 = ((y^0)^2 - (y^1)^2 - (y^2)^2 - (y^3)^2)\vec{f}_0\end{aligned}.$$

If we want these equalities to be right the conditions

$$\begin{array}{llll} c_{00} = 1 & c_{01} + c_{10} = 0 & c_{12} + c_{21} = 0 & c_{23} + c_{32} = 0 \\ c_{11} = -1 & c_{02} + c_{20} = 0 & c_{13} + c_{31} = 0 & \\ c_{22} = -1 & c_{03} + c_{30} = 0 & & \\ c_{33} = -1 & & & \end{array}$$

are to be carried out.
If to express $c_{ij}$ through $a_i^j$ last conditions give:

$$\begin{aligned}(a_0^0)^2 - (a_0^1)^2 - (a_0^2)^2 - (a_0^3)^2 &= 1 \\ (a_1^0)^2 - (a_1^1)^2 - (a_1^2)^2 - (a_1^3)^2 &= -1 \\ (a_2^0)^2 - (a_2^1)^2 - (a_2^2)^2 - (a_2^3)^2 &= -1 \\ (a_3^0)^2 - (a_3^1)^2 - (a_3^2)^2 - (a_3^3)^2 &= -1\end{aligned}$$

$$\begin{aligned}a_0^0 a_1^0 - a_0^1 a_1^1 - a_0^2 a_1^2 - a_0^3 a_1^3 &= 0 \\ a_0^0 a_2^0 - a_0^1 a_2^1 - a_0^2 a_2^2 - a_0^3 a_2^3 &= 0 \\ a_0^0 a_3^0 - a_0^1 a_3^1 - a_0^2 a_3^2 - a_0^3 a_3^3 &= 0 \\ a_1^0 a_2^0 - a_1^1 a_2^1 - a_1^2 a_2^2 - a_1^3 a_2^3 &= 0 \\ a_1^0 a_3^0 - a_1^1 a_3^1 - a_1^2 a_3^2 - a_1^3 a_3^3 &= 0 \\ a_2^0 a_3^0 - a_2^1 a_3^1 - a_2^2 a_3^2 - a_2^3 a_3^3 &= 0\end{aligned}.$$

These are Lorentz's transformations. So, the **Theorem 3.1** is proved: linear transformation of basis and the coordinates, keeping a kind (1.1) for product of a vector and its conjugated, is Lorentz's transformation.
Let's consider a special case of Lorentz's transformation, namely, turn of spatial axes:

$$(a_i^j) = \begin{pmatrix} 1 & 0 & 0 & 0 \\ 0 & a_1^1 & a_1^2 & a_1^3 \\ 0 & a_2^1 & a_2^2 & a_2^3 \\ 0 & a_3^1 & a_3^2 & a_3^3 \end{pmatrix} .$$

Directing cosines of new axes submit to conditions:

$$\begin{aligned} &(a_1^1)^2 + (a_1^2)^2 + (a_1^3)^2 = 1 \\ &(a_2^1)^2 + (a_2^2)^2 + (a_2^3)^2 = 1 \\ &(a_3^1)^2 + (a_3^2)^2 + (a_3^3)^2 = 1 \\ &a_1^1 a_2^1 + a_1^2 a_2^2 + a_1^3 a_2^3 = 0 \\ &a_1^1 a_3^1 + a_1^2 a_3^2 + a_1^3 a_3^3 = 0 \\ &a_2^1 a_3^1 + a_2^2 a_3^2 + a_2^3 a_3^3 = 0 \end{aligned} . \tag{3.1}$$

Now we shall find out, how products $\vec{k}_i \vec{k}_j$ in the form of linear combinations of the same vectors are expressed:

$$\begin{aligned} \vec{k}_i \vec{k}_j &= (\sum_m a_i^m \vec{e}_m)(\sum_n a_j^n \vec{e}_n) = \\ &= \sum_{m,n} a_i^m a_j^n \vec{e}_m \vec{e}_n \end{aligned} \tag{3.2}$$

System (3.1) has a lot of solutions. We solve it, for example, concerning variables $a_1^3, a_2^2, a_2^3, a_3^1, a_3^2, a_3^3$, i.e. we express these last ones through the remained variables $a_1^1, a_1^2, a_2^1$. So to within 3 arbitrary elements matrix $(a_i^j)$ of transformation of coordinates is found. Further we find the matrix $(b_i^j)$ of transformation of basis opposite to it. Then in equality (3.2) we replace products $\vec{e}_m \vec{e}_n$, using parities (1.11). But the right parts of parities (1.11) are proportional to the same vector $\vec{R}$ which is a linear combination of basic vectors $\vec{e}_i$. Using a matrix $(b_i^j)$ of transformation of basis, we obtain value of a vector $\vec{R}$, and consequently the right parts of expressions (3.2) in the form of linear combinations of new basic vectors $\vec{k}_i$. It is clear, that factors in these linear combinations depend on 3 arbitrary parameters $a_1^1, a_1^2, a_2^1$.
We choose these arbitrary parameters in the special form; for example, so that directing cosines of the first spatial axis appeared to be equal to parameters of the constructed algebra $\alpha^i$:

$$\begin{aligned} a_1^1 &= \alpha^1 \\ a_1^2 &= \alpha^2 \\ a_1^3 &= \alpha^3 \end{aligned} .$$

Then in new coordinate system the algebra gets especially simple kind:

$$
\begin{array}{ll}
\vec{k}_0\vec{k}_0 = a(\vec{k}_0+\vec{k}_1) & \vec{k}_2\vec{k}_0 = 0 \\
\vec{k}_0\vec{k}_1 = -a(\vec{k}_0+\vec{k}_1) & \vec{k}_2\vec{k}_1 = 0 \\
\vec{k}_0\vec{k}_2 = 0 & \vec{k}_2\vec{k}_2 = a(\vec{k}_0+\vec{k}_1) \\
\vec{k}_0\vec{k}_3 = 0 & \vec{k}_2\vec{k}_3 = c(\vec{k}_0+\vec{k}_1) \\
\vec{k}_1\vec{k}_0 = -a(\vec{k}_0+\vec{k}_1) & \vec{k}_3\vec{k}_0 = 0 \\
\vec{k}_1\vec{k}_1 = a(\vec{k}_0+\vec{k}_1) & \vec{k}_3\vec{k}_1 = 0 \\
\vec{k}_1\vec{k}_2 = 0 & \vec{k}_3\vec{k}_2 = -c(\vec{k}_0+\vec{k}_1) \\
\vec{k}_1\vec{k}_3 = 0 & \vec{k}_3\vec{k}_3 = a(\vec{k}_0+\vec{k}_1)
\end{array}
\qquad (3.3)
$$

Thus, the **Theorem 3.2** is proved: if to choose new system of coordinates in the next special image, i.e. to direct a basic vector $\vec{k}_1$ of new system so that its directing cosines in previous system have coincided with parameters $\alpha^i$ of algebra, then rules of multiplication of basic vectors follow parities (3.3).

**The remark**. With the help of a corresponding choice of system of coordinates it is always possible to achieve the fact that two of three factors $\alpha^i$ in all algebraic and analytical parities became equal to zero. The third factor with necessity becomes equal to unity.

### 4. Algebra and Analysis in pseudo-Riemannian Space.

We shall pass to curvilinear pseudo-Riemannian space with the metric $g_{ij}(x^0,x^1,x^2,x^3)$, satisfying the conditions:

$$
\begin{array}{l}
g_{ij} = g_{ji} \\
g_{00} > 0 \\
\begin{vmatrix} g_{00} & g_{01} \\ g_{10} & g_{11} \end{vmatrix} < 0 \\
\begin{vmatrix} g_{00} & g_{01} & g_{02} \\ g_{10} & g_{11} & g_{12} \\ g_{20} & g_{21} & g_{22} \end{vmatrix} > 0
\end{array}
\qquad (4.1)
$$

$$
\begin{vmatrix} g_{00} & g_{01} & g_{02} & g_{03} \\ g_{10} & g_{11} & g_{12} & g_{13} \\ g_{20} & g_{21} & g_{22} & g_{23} \\ g_{30} & g_{31} & g_{32} & g_{33} \end{vmatrix} < 0 \qquad (4.1)
$$

As it is known [3, p. 301], metric tensor in any frame of reference, realizable by means of real bodies, should satisfy these conditions.

We shall fix a point $(x_0^0, x_0^1, x_0^2, x_0^3)$ and the metric $g_{ij}(x_0^0, x_0^1, x_0^2, x_0^3) = g_{ij}$ in it.

Let's consider pseudoeuclidean space, tangent to initial curvilinear space in this point. We shall enter in all tangent space the metric $g_{ij}$. Owing to conditions (4.1) there is a unique triangular transformation of the basis $\{\vec{e}_i\}$, leading the quadratic form $\sum_{i,j=0}^{3} g_{ij}x^i x^j$ to a normal kind [2, pp. 195-199]

$$(y^0)^2-(y^1)^2-(y^2)^2-(y^3)^2 \ .$$

Further, where it will not cause bewilderment, we shall omit summation limits.
We shall find a matrix

$$A=(a_i^j)=\begin{bmatrix} a_0^0 & 0 & 0 & 0 \\ a_1^0 & a_1^1 & 0 & 0 \\ a_2^0 & a_2^1 & a_2^2 & 0 \\ a_3^0 & a_3^1 & a_3^2 & a_3^3 \end{bmatrix}$$

of the above-stated triangular transformation $\vec{k}_i=\sum_j a_i^j \vec{e}_j$. Instead of use of Jakoby's method [2, p. 195-199] of matrix $A$ determination we shall take advantage of that during transformation by means of matrix $A$ of a basis $\{\vec{e}_i\}$ to basis $\{\vec{k}_i\}$ the matrix $G=(g_{ij})$ will be transformed under the law $AGA^T$ where $A^T$ is the transposed matrix $A$ [2, p. 189]. I.e. we shall solve system of the equations

$$AGA^T=J \qquad \text{or} \qquad \sum_{m,n} a_i^m g_{mn}(a^T)_n^j=\sum_{m,n} g_{mn}a_i^m a_j^n=J_{ij} \quad ,$$

where

$$J=\begin{pmatrix} 1 & 0 & 0 & 0 \\ 0 & -1 & 0 & 0 \\ 0 & 0 & -1 & 0 \\ 0 & 0 & 0 & -1 \end{pmatrix}.$$

This system of the nonlinear equations for variables $a_j^i$ because of a specific kind of a matrix $A$ can be solved easily by means of consecutive exception of groups of variables $(a_0^0,a_1^0,a_2^0,a_3^0),(a_1^1,a_2^1,a_3^1),(a_2^2,a_3^2),a_3^3$. As obvious expressions for $a_j^i$ are too bulky, we shall not result them here. Further let $\vec{e}_i=\sum_j b_i^j \vec{k}_j$, i.e $B=(b_j^i)=A^{-1}$.

If we enter in tangent space algebra with rules of multiplication (1.11) of basic vectors $\vec{k}_i$ where $a,c,\alpha^0,\alpha^1,\alpha^2,\alpha^3$ are the real constants $(\alpha^0=1,\quad \alpha^{1^2}+\alpha^{2^2}+\alpha^{3^2}=1), \vec{R}=\alpha^0\vec{k}_0+\alpha^1\vec{k}_1+\alpha^2\vec{k}_2+\alpha^3\vec{k}_3$, then, as we know matrices $(a_i^j)$ and $(b_i^j)$, it is possible to find rules of multiplication of basic vectors $\vec{e}_i$. The law of multiplication (1.11) of basic vectors $\vec{k}_i$ in these designations is:

$$\vec{k}_i\vec{k}_j = \kappa_{ij}\vec{R} = \kappa_{ij}\sum_m \alpha^m \vec{k}_m \ ,$$

where

$$(\kappa_{ij}) = \begin{pmatrix} a & -\alpha^1 a & -\alpha^2 a & -\alpha^3 a \\ -\alpha^1 a & a & \alpha^3 c & -\alpha^2 c \\ -\alpha^2 a & -\alpha^3 c & a & \alpha^1 c \\ -\alpha^3 a & \alpha^2 c & -\alpha^1 c & a \end{pmatrix} .$$

Then

$$\begin{aligned} \vec{e}_i\vec{e}_j &= (\sum_m b_i^m \vec{k}_m)(\sum_m b_j^n \vec{k}_n) = \sum_{m,n} b_i^m b_j^n (\vec{k}_m \vec{k}_n) = \\ &= \sum_{m,n} b_i^m b_j^n \kappa_{mn} (\sum_t \alpha^t \vec{k}_t) = h_{ij} (\sum_t \alpha^t \vec{k}_t) \end{aligned} ,$$

where

$$h_{ij} = \sum_{m,n} b_i^m b_j^n \kappa_{mn} \quad . \tag{4.2}$$

And further

$$\begin{aligned} \vec{e}_i\vec{e}_j &= h_{ij}(\sum_m \alpha^m \vec{k}_m) = h_{ij}\sum_m \alpha^m (\sum_n a_m^n \vec{e}_n) = \\ &= h_{ij}\sum_n (\sum_m \alpha^m a_m^n)\vec{e}_n = h_{ij}\sum_n \vartheta^n \vec{e}_n = h_{ij}\vec{R} \end{aligned} , \tag{4.3}$$

where

$$\vec{R} = \sum_n \vartheta^n \vec{e}_n \tag{4.4}$$

and

$$\vartheta^i = \sum_j \alpha^j a_j^i \quad . \tag{4.5}$$

So, rules of multiplication of basic vectors $\vec{e}_i$ are given by formulas (4.2-5).

Everything that has been told earlier about division of elements in pseudoeuclidean space, remains correct in pseudo-Riemannian one. We shall enter in all curvilinear space local geometry

$$\begin{aligned} d\vec{r}(x^0,x^1,x^2,x^3) &= dx^0\vec{e}_0(x^0,x^1,x^2,x^3) + dx^1\vec{e}_1(x^0,x^1,x^2,x^3) + \\ &+ dx^2\vec{e}_2(x^0,x^1,x^2,x^3) + dx^3\vec{e}_3(x^0,x^1,x^2,x^3) \end{aligned} .$$

The local algebra is entered by introduction of operation of multiplication of basic vectors $\vec{e}_i$:

$$\vec{e}_i(x^0,x^1,x^2,x^3)\vec{e}_j(x^0,x^1,x^2,x^3)=h_{ij}(x^0,x^1,x^2,x^3;\alpha^1,\alpha^2,\alpha^3,a,c)\vec{R} \ ,$$

where $\vec{R}=\sum_i \vartheta^i(x^0,x^1,x^2,x^3;\alpha^1,\alpha^2,\alpha^3,a,c)\vec{e}_i(x^0,x^1,x^2,x^3)$.

Thus it is necessary to emphasize that $\alpha^1,\alpha^2,\alpha^3,a,c$ are the real constants, uniform for all pseudo-Riemannian space.
We shall enter on variety $\{x^i\}$ variety of functions

$$\vec{u}(\vec{r})=\{u^0(x^0,x^1,x^2,x^3),u^1(x^0,x^1,x^2,x^3),u^2(x^0,x^1,x^2,x^3),u^3(x^0,x^1,x^2,x^3)\} \ ,$$

and on this variety the same local geometry and algebra are entered:

$$\begin{aligned} D\vec{u}(x^0,x^1,x^2,x^3) = {} & Du^0(x^0,x^1,x^2,x^3)\vec{e}_0(x^0,x^1,x^2,x^3)+ \\ & +Du^1(x^0,x^1,x^2,x^3)\vec{e}_1(x^0,x^1,x^2,x^3)+ \\ & +Du^2(x^0,x^1,x^2,x^3)\vec{e}_2(x^0,x^1,x^2,x^3)+ \\ & +Du^3(x^0,x^1,x^2,x^3)\vec{e}_3(x^0,x^1,x^2,x^3) \end{aligned} \tag{4.6}$$

where

$$\begin{aligned} Du^i(x^0,x^1,x^2,x^3) &= \sum_j [u^i_{,x^j}(x^0,x^1,x^2,x^3)+\sum_k \Gamma^i_{kj}(x^0,x^1,x^2,x^3)u^k(x^0,x^1,x^2,x^3)]dx^j = \\ &= \sum_j u^i_{;x^j}(x^0,x^1,x^2,x^3)dx^j \end{aligned} \ . \tag{4.7}$$

Here $\Gamma^i_{kj}$ are Christoffel's symbols, and $u^i_{;x^j}$ are covariant derivatives of vector $\vec{u}$. We name the left derivative of function $\vec{u}(\vec{r})$ in a point $\vec{r}$ not dependent on a way for $d\vec{r}$ to reach zero a limit (if it, certainly, exists) of difference expression $[\frac{D\vec{u}(\vec{r})}{d\vec{r}}]_l$:

$$\vec{u}'_l(\vec{r})=\lim_{d\vec{r}\to 0}[\frac{D\vec{u}(\vec{r})}{d\vec{r}}]_l \ .$$

We shall enter designations $\vec{u}'_l(\vec{r})=v^0\vec{e}_0+v^1\vec{e}_1+v^2\vec{e}_2+v^3\vec{e}_3$. Considering parities (4.8) and omitting arguments of $Du^i(x^0,x^1,x^2,x^3)$, we shall obtain:

$$\begin{aligned} \vec{u}'_l(\vec{r}) &= \lim_{d\vec{r}\to 0}[\frac{D\vec{u}(\vec{r})}{d\vec{r}}]_l = \\ &= \lim_{\substack{dx^0\to 0\\ dx^1\to 0\\ dx^2\to 0\\ dx^3\to 0}} \frac{Du^0\vec{e}_0+Du^1\vec{e}_1+Du^2\vec{e}_2+Du^3\vec{e}_3}{dx^0\vec{e}_0+dx^1\vec{e}_1+dx^2\vec{e}_2+dx^3\vec{e}_3} = \end{aligned}$$

$$= \lim_{\substack{dx^0 \to 0 \\ dx^1 \to 0 \\ dx^2 \to 0 \\ dx^3 \to 0}} \{[(u^0_{;x^0} dx^0 + u^0_{;x^1} dx^1 + u^0_{;x^2} dx^2 + u^0_{;x^3} dx^3)\vec{e}_0 +$$

$$+ (u^1_{;x^0} dx^0 + u^1_{;x^1} dx^1 + u^1_{;x^2} dx^2 + u^1_{;x^3} dx^3)\vec{e}_1 +$$
$$+ (u^2_{;x^0} dx^0 + u^2_{;x^1} dx^1 + u^2_{;x^2} dx^2 + u^2_{;x^3} dx^3)\vec{e}_2 + \quad .$$
$$+ (u^3_{;x^0} dx^0 + u^3_{;x^1} dx^1 + u^3_{;x^2} dx^2 + u^3_{;x^3} dx^3)\vec{e}_3]/$$
$$/[dx^0\vec{e}_0 + dx^1\vec{e}_1 + dx^2\vec{e}_2 + dx^3\vec{e}_3]\}$$

Again, as like as in a case of pseudoeuclidean space, we shall assume $d\vec{r} = dx^0\vec{e}_0$. Then

$$\vec{u}_l'(\vec{r}) = \frac{u^0_{;x^0}\vec{e}_0 + u^1_{;x^0}\vec{e}_1 + u^2_{;x^0}\vec{e}_2 + u^3_{;x^0}\vec{e}_3}{\vec{e}_0} = \frac{A\vec{R}}{\vec{e}_0} =$$
$$= \frac{A\vartheta^0\vec{e}_0 + A\vartheta^1\vec{e}_1 + A\vartheta^2\vec{e}_2 + A\vartheta^3\vec{e}_3}{\vec{e}_0} \quad .$$

From here we obtain

$$u^0_{;x^0} = A\vartheta^0$$
$$u^1_{;x^0} = A\vartheta^1$$
$$u^2_{;x^0} = A\vartheta^2 \quad .$$
$$u^3_{;x^0} = A\vartheta^3$$

And at last analogues of conditions of Cauchy-Riemann are:

$$u^1_{;x^0} = \frac{\vartheta^1}{\vartheta^0} u^0_{;x^0}$$
$$u^2_{;x^0} = \frac{\vartheta^2}{\vartheta^0} u^0_{;x^0} \qquad (4.8)$$
$$u^3_{;x^0} = \frac{\vartheta^3}{\vartheta^0} u^0_{;x^0}$$

Analogues of conditions of Cauchy-Riemann for $d\vec{r} = dx^i\vec{e}_i$ , $i = 1,2,3$ turn out similarly. If now

$$d\vec{r} = dx^0\vec{e}_0 + dx^1\vec{e}_1 + dx^2\vec{e}_2 + dx^3\vec{e}_3 = d\tau\vec{R} =$$
$$= d\tau\vartheta^0\vec{e}_0 + d\tau\vartheta^1\vec{e}_1 + d\tau\vartheta^2\vec{e}_2 + d\tau\vartheta^3\vec{e}_3 \quad ,$$

then

$$\begin{aligned} dx^0 &= \vartheta^0 d\tau \\ dx^1 &= \vartheta^1 d\tau \\ dx^2 &= \vartheta^2 d\tau \\ dx^3 &= \vartheta^3 d\tau \end{aligned} \quad . \qquad (4.8^1)$$

And further

$$\begin{aligned} \vec{u}'(\vec{r}) = \lim_{B\to 0}\{[(&u^0_{;x^0} d\tau\vartheta^0 + u^0_{;x^1} d\tau\vartheta^1 + u^0_{;x^2} d\tau\vartheta^2 + u^0_{;x^3} d\tau\vartheta^3)\vec{e}_0 + \\ &+(u^1_{;x^0} d\tau\vartheta^0 + u^1_{;x^1} d\tau\vartheta^1 + u^1_{;x^2} d\tau\vartheta^2 + u^1_{;x^3} d\tau\vartheta^3)\vec{e}_1 + \\ &+(u^2_{;x^0} d\tau\vartheta^0 + u^2_{;x^1} d\tau\vartheta^1 + u^2_{;x^2} d\tau\vartheta^2 + u^2_{;x^3} d\tau\vartheta^3)\vec{e}_2 + \\ &+(u^3_{;x^0} d\tau\vartheta^0 + u^3_{;x^1} d\tau\vartheta^1 + u^3_{;x^2} d\tau\vartheta^2 + u^3_{;x^3} d\tau\vartheta^3)\vec{e}_3]/ \\ &/[d\tau(\vartheta^0\vec{e}_0 + \vartheta^1\vec{e}_1 + \vartheta^2\vec{e}_2 + \vartheta^3\vec{e}_3)]\} \end{aligned} \quad ,$$

whence $u^i_{;x^0}\vartheta^0 + u^i_{;x^1}\vartheta^1 + u^i_{;x^2}\vartheta^2 + u^i_{;x^3}\vartheta^3 = 0$ can be obtained.

And at last the **Theorem 4.1** is proved: analogues of conditions of Cauchy-Riemann for pseudo-Riemannian space are given by parities:

$$\begin{aligned} \vartheta^0 u^0_{;x^0} + \vartheta^1 u^0_{;x^1} + \vartheta^2 u^0_{;x^2} + \vartheta^3 u^0_{;x^3} &= 0 \\ \vartheta^0 u^1_{;x^0} + \vartheta^1 u^1_{;x^1} + \vartheta^2 u^1_{;x^2} + \vartheta^3 u^1_{;x^3} &= 0 \\ \vartheta^0 u^2_{;x^0} + \vartheta^1 u^2_{;x^1} + \vartheta^2 u^2_{;x^2} + \vartheta^3 u^2_{;x^3} &= 0 \\ \vartheta^0 u^3_{;x^0} + \vartheta^1 u^3_{;x^1} + \vartheta^2 u^3_{;x^2} + \vartheta^3 u^3_{;x^3} &= 0 \end{aligned} \qquad (4.9)$$

$$\begin{aligned} u^1_{;x^0} &= \frac{\vartheta^1}{\vartheta^0} u^0_{;x^0} & u^1_{;x^1} &= \frac{\vartheta^1}{\vartheta^0} u^0_{;x^1} \\ u^2_{;x^0} &= \frac{\vartheta^2}{\vartheta^0} u^0_{;x^0} & u^2_{;x^1} &= \frac{\vartheta^2}{\vartheta^0} u^0_{;x^1} \\ u^3_{;x^0} &= \frac{\vartheta^3}{\vartheta^0} u^0_{;x^0} & u^3_{;x^1} &= \frac{\vartheta^3}{\vartheta^0} u^0_{;x^1} \\ u^1_{;x^2} &= \frac{\vartheta^1}{\vartheta^0} u^0_{;x^2} & u^1_{;x^3} &= \frac{\vartheta^1}{\vartheta^0} u^0_{;x^3} \\ u^2_{;x^2} &= \frac{\vartheta^2}{\vartheta^0} u^0_{;x^2} & u^2_{;x^3} &= \frac{\vartheta^2}{\vartheta^0} u^0_{;x^3} \\ u^3_{;x^2} &= \frac{\vartheta^3}{\vartheta^0} u^0_{;x^2} & u^3_{;x^3} &= \frac{\vartheta^3}{\vartheta^0} u^0_{;x^3} \end{aligned} \qquad (4.9)$$

**The remark**. If any of first four equations (4.9) is multiplied by $d\tau$, then considering parities ($4.8^1$) and (4.7) we shall obtain

$$\begin{aligned} &d\tau\,\vartheta^0 u^i_{;x^0} + d\tau\,\vartheta^1 u^i_{;x^1} + d\tau\,\vartheta^2 u^i_{;x^2} + d\tau\,\vartheta^3 u^i_{;x^3} = \\ &= dx^0 u^i_{;x^0} + dx^1 u^i_{;x^1} + dx^2 u^i_{;x^2} + dx^3 u^i_{;x^3} = \\ &= Du^i = 0 \end{aligned} \quad , \qquad (4.10)$$

i.e. vector $\vec{u}$ is transferred in parallel (is constant) along the curve ($4.8^1$). Besides, by direct calculations it is possible to be convinced that if to preserve the dependences ($4.8^1$) between all $dx^i$ the parity

$$\begin{aligned}
\sum_{i,j} g_{ij} dx^i dx^j &= \sum_{i,j} g_{ij} (d\tau \vartheta^i)(d\tau \vartheta^j) = d\tau^2 \sum_{i,j} g_{ij} \vartheta^i \vartheta^j = \\
&= d\tau^2 \sum_{i,j} g_{ij} (\sum_m \alpha^m a_m^i)(\sum_n \alpha^n a_n^j) = \\
&= d\tau^2 \sum_{i,j,m,n} g_{ij} \alpha^m \alpha^n a_m^i a_n^j = \\
&= d\tau^2 \sum_{m,n} \alpha^m \alpha^n (\sum_{i,j} g_{ij} a_m^i a_n^j) = \\
&= d\tau^2 \sum_{m,n} \alpha^m \alpha^n J_{mn} = \\
&= d\tau^2 (1 - (\alpha_1)^2 - (\alpha_2)^2 - (\alpha_2)^2) = 0
\end{aligned} \tag{4.11}$$

is carried out, i.e. that the curve ($4.8^1$) is zero. The question when this curve is geodetic deserves the examination in an individual paragraph.

## 5. The Central-symmetric Case.

We shall enter the following designations:

$$x^0 = t, \quad x^1 = r, \quad x^2 = \vartheta, \quad x^3 = \varphi \ .$$

Then the most general central-symmetric expression for $ds^2$ [3, p. 382] is:

$$ds^2 = -h(r,t)dr^2 - r^2(d\vartheta^2 + \sin^2\vartheta d\varphi^2) + l(r,t)dt^2 \ ,$$

i.e.

$$g_{00} = l, \quad g_{11} = -h, \quad g_{22} = -r^2, \quad g_{33} = -r^2 \sin^2\vartheta$$

and, accordingly

$$g^{00} = l^{-1}, \quad g^{11} = -h^{-1}, \quad g^{22} = -r^{-2}, \quad g^{33} = -r^{-2}\sin^{-2}\vartheta \ .$$

Let's designate differentiation along "time" $t$ by the top point, and differentiation along "radius" $r$ by the right top stroke.
Then symbols of Christoffel accept values:

$$\Gamma^0_{00} = \frac{\dot{l}}{2l}, \quad \Gamma^0_{01} = \Gamma^0_{10} = \frac{l'}{2l}, \quad \Gamma^0_{02} = \Gamma^0_{20} = 0, \quad \Gamma^0_{03} = \Gamma^0_{30} = 0,$$
$$\Gamma^0_{11} = \frac{\dot{h}}{2l}, \quad \Gamma^0_{12} = \Gamma^0_{21} = 0, \quad \Gamma^0_{13} = \Gamma^0_{31} = 0,$$
$$\Gamma^0_{22} = 0, \quad \Gamma^0_{23} = \Gamma^0_{32} = 0,$$
$$\Gamma^0_{33} = 0,$$
$$\Gamma^1_{00} = \frac{l'}{2h}, \quad \Gamma^1_{01} = \Gamma^1_{10} = \frac{\dot{h}}{2h}, \quad \Gamma^1_{02} = \Gamma^1_{20} = 0, \quad \Gamma^1_{03} = \Gamma^1_{30} = 0,$$
$$\Gamma^1_{11} = \frac{h'}{2h}, \quad \Gamma^1_{12} = \Gamma^1_{21} = 0, \quad \Gamma^1_{13} = \Gamma^1_{31} = 0,$$
$$\Gamma^1_{22} = -\frac{r}{h}, \quad \Gamma^1_{23} = \Gamma^1_{32} = 0,$$
$$\Gamma^1_{33} = -\frac{r \sin^2 \vartheta}{h},$$
$$\Gamma^2_{00} = 0, \quad \Gamma^2_{01} = \Gamma^2_{10} = 0, \quad \Gamma^2_{02} = \Gamma^2_{20} = 0, \quad \Gamma^2_{03} = \Gamma^2_{30} = 0,$$
$$\Gamma^2_{11} = 0, \quad \Gamma^2_{12} = \Gamma^2_{21} = \frac{1}{r}, \quad \Gamma^2_{13} = \Gamma^2_{31} = 0,$$
$$\Gamma^2_{22} = 0, \quad \Gamma^2_{23} = \Gamma^2_{32} = 0,$$
$$\Gamma^2_{33} = -\sin\vartheta \cos\vartheta,$$
$$\Gamma^3_{00} = 0, \quad \Gamma^3_{01} = \Gamma^3_{10} = 0, \quad \Gamma^3_{02} = \Gamma^3_{20} = 0, \quad \Gamma^3_{03} = \Gamma^3_{30} = 0, \qquad .$$
$$\Gamma^3_{11} = 0, \quad \Gamma^3_{12} = \Gamma^3_{21} = 0, \quad \Gamma^3_{13} = \Gamma^3_{31} = \frac{1}{r},$$
$$\Gamma^3_{22} = 0, \quad \Gamma^3_{23} = \Gamma^3_{32} = \frac{\cos\vartheta}{\sin\vartheta},$$
$$\Gamma^3_{33} = 0$$

As algebra parameters $\alpha^i$ should satisfy only one parity (1.12), we shall choose them as follows: $\alpha^1 = 1, \alpha^2 = 0, \alpha^3 = 0$ . From here, considering parities (4.5), we shall obtain: $\vartheta^0 = \frac{1}{\sqrt{l}}$, $\vartheta^1 = -\frac{1}{\sqrt{h}}$, $\vartheta^2 = \vartheta^3 = 0$. Now we shall enter designations:
$u^0 = \tau, \quad u^1 = u, \quad u^0 = v, \quad u^0 = w$ .
From here we shall obtain values of covariant derivatives of functions $u^i$:

$$\begin{aligned} u^0_{;x^0} &= u^0_{x^0} + \Gamma^0_{00} u^0 + \Gamma^0_{10} u^1 + \Gamma^0_{20} u^2 + \Gamma^0_{30} u^3 = \\ &= u^0_{x^0} + \frac{\dot{l}}{2l} u^0 + \frac{l'}{2l} u^1 = \\ &= \dot{\tau} + \frac{\dot{l}}{2l} \tau + \frac{l'}{2l} u \end{aligned} \quad ,$$

and further

$$u^0_{;x^1} = \tau' + \frac{l'}{2l}\tau + \frac{\dot{h}}{2l}u$$
$$u^0_{;x^2} = \tau_\vartheta$$
$$u^0_{;x^3} = \tau_\varphi$$
$$u^1_{;x^0} = \dot{u} + \frac{l'}{2h}\tau + \frac{\dot{h}}{2h}u$$
$$u^1_{;x^1} = u' + \frac{\dot{h}}{2h}\tau + \frac{h'}{2h}u$$
$$u^1_{;x^2} = u_\vartheta - \frac{r}{h}v$$
$$u^1_{;x^3} = u_\varphi - \frac{r\sin^2\vartheta}{h}w$$
$$u^2_{;x^0} = \dot{v}$$
$$u^2_{;x^1} = v' + \frac{1}{r}v$$
$$u^2_{;x^2} = v_\vartheta + \frac{1}{r}u$$
$$u^2_{;x^3} = v_\varphi - \sin\vartheta\cos\vartheta\, w$$
$$u^3_{;x^0} = \dot{w}$$
$$u^3_{;x^1} = w' + \frac{1}{r}w$$
$$u^3_{;x^2} = w_\vartheta + \frac{\cos\vartheta}{\sin\theta}w$$
$$u^3_{;x^3} = w_\varphi + \frac{1}{r}u + \frac{\cos\vartheta}{\sin\theta}v$$

If to consider all above-stated the system (4.9) gets a form:

$$\frac{1}{2\sqrt{h}\,l^{\frac{3}{2}}}(-\sqrt{l}\,(\tau\, l' + u\,\dot{h} + 2\,l\,\tau') + \sqrt{h}\,(u\,l' + \tau\,\dot{l} + 2\,l\,\dot{\tau})) = 0 \tag{5.1}$$
$$\frac{1}{2\,h^{\frac{3}{2}}\sqrt{l}}(-\sqrt{l}(u\,h' + \tau\,\dot{h} + 2\,h u') + \sqrt{h}(\tau\, l' + u\,\dot{h} + 2\,h\dot{u})) = 0 \tag{5.2}$$
$$-\frac{\frac{v}{r} + v'}{\sqrt{h}} + \frac{\dot{v}}{\sqrt{l}} = 0 \tag{5.3}$$
$$-\frac{\frac{w}{r} + w'}{\sqrt{h}} + \frac{\dot{w}}{\sqrt{l}} = 0 \tag{5.4}$$
$$\frac{1}{2}(\frac{\tau\, l' + u\,\dot{h}}{h} + 2\,\dot{u} + \frac{u\,l' + \tau\,\dot{l} + 2\,l\,\dot{\tau}}{\sqrt{h\,l}}) = 0 \tag{5.5}$$
$$\dot{v} = 0 \tag{5.6}$$
$$\dot{w} = 0 \tag{5.7}$$

$$\frac{1}{2}\left(\frac{u\,h'+\tau\,\dot{h}}{h}+2u'+\frac{\tau\,l'+u\,\dot{h}+2\,l\,\tau'}{\sqrt{h\,l}}\right)=0 \tag{5.8}$$

$$\frac{v}{r}+v'=0 \tag{5.9}$$

$$\frac{w}{r}+w'=0 \tag{5.10}$$

$$-\frac{r\,v}{h}+u_\vartheta+\frac{\sqrt{l}\,\tau_\vartheta}{\sqrt{h}}=0 \tag{5.11}$$

$$\frac{u}{r}+v_\vartheta=0 \tag{5.12}$$

$$ctg\vartheta\, w+w_\vartheta=0 \tag{5.13}$$

$$-\frac{r\sin^2\vartheta\, w}{h}+u_\varphi+\frac{\sqrt{l}\,\tau_\varphi}{\sqrt{h}}=0 \tag{5.14}$$

$$-\cos\vartheta\sin\vartheta\, w+v_\varphi=0 \tag{5.15}$$

$$\frac{u}{r}+ctg\vartheta\, v+w_\varphi=0 \tag{5.16}$$

From the equations (5.6) and (5.7)

$v=v(r,\vartheta,\varphi)$ and $w=w(r,\vartheta,\varphi)$

immediately follows.
We substitute these values in (5.9) and (5.10), and, solving the obtained equations, we get:

$v=\dfrac{\tilde{v}(\vartheta,\varphi)}{r}$ and $w=\dfrac{\tilde{w}(\vartheta,\varphi)}{r}$ .

We substitute last value in the equation (5.13) and we specify a function $w$ kind:

$$w=\frac{\tilde{w}(\varphi)}{r\sin\vartheta}\ . \tag{5.17}$$

Further, after substitution of last values of functions $v$ and $w$ in the equations (5.12) and (5.16), we obtain a new kind of these equations:

$$\frac{u+\tilde{v}_\vartheta}{r}=0 \tag{5.12a}$$

$$\frac{u+ctg\vartheta\,\tilde{v}+\csc\vartheta\ \tilde{w}_\varphi}{r}=0 \tag{5.16a}$$

Getting rid of function $u$, we obtain the equation

$$ctg\vartheta\,\tilde{v}+\csc\vartheta\,\tilde{w}_\varphi-\tilde{v}_\vartheta=0\ .$$

While solving this equation for function $\tilde{v}$ we obtain:

$$\tilde{v} = \sin\vartheta\,\tilde{\tilde{v}}(\varphi) - \cos\vartheta\,\tilde{w}_{\varphi}(\varphi)$$

or

$$v = \frac{\sin\vartheta\,\tilde{\tilde{v}}(\varphi) - \cos\vartheta\,\tilde{w}_{\varphi}(\varphi)}{r} \quad . \tag{5.18}$$

After substitution of values of functions $v$ and $w$ accordingly from formulas (5.17) and (5.18) in the equation (5.15) the last one takes a form:

$$-\frac{\cos\vartheta\,\tilde{w}(\varphi) - \sin\vartheta\,\tilde{\tilde{v}}_{\varphi}(\varphi) + \cos\vartheta\,\tilde{w}_{\varphi\varphi}(\varphi)}{r} = 0 \quad . \tag{5.15a}$$

We solve the obtained equation for function $\tilde{\tilde{v}}(\varphi)$:

$$\tilde{\tilde{v}}(\varphi) = V + ctg\vartheta \int (\tilde{w}(\varphi) + \tilde{w}_{\varphi\varphi}(\varphi))\,d\varphi \quad ,$$

where $V = const$ .

We shall enter a designation $W(\varphi) = \int \tilde{w}(\varphi)\,d\varphi$. Then we obtain:

$$\tilde{\tilde{v}}(\varphi) = V + ctg\vartheta(W(\varphi) + W_{\varphi\varphi}(\varphi)) \ ,$$

and accordingly to the expression (5.18)

$$v = \frac{\sin\vartheta(V + ctg\vartheta(W(\varphi) + W_{\varphi\varphi}(\varphi))) - \cos(\vartheta)W_{\varphi\varphi}(\varphi)}{r} \quad . \tag{5.18a}$$

Expression (5.17) gets a form:

$$w = \frac{W_{\varphi}(\varphi)}{r\sin\vartheta} \quad . \tag{5.17a}$$

As a result the equations (5.12a) and (5.16a) turn to the equations

$$\frac{V\cos\vartheta + u - \sin\vartheta W(\varphi)}{r} = 0 \quad , \tag{5.12b}$$

$$\frac{V\cos\vartheta + u + \cos\vartheta\,ctg\vartheta W(\varphi) + \csc\vartheta W_{\varphi\varphi}(\varphi)}{r} = 0 \quad . \tag{5.16b}$$

Again getting rid of $u$ in the obtained system of the equations, we obtain the equation

$$(\cos\vartheta\,ctg\vartheta + \sin\vartheta)\ W(\varphi) + \csc\vartheta\ W_{\varphi\varphi}(\varphi) = 0 \ .$$

Integration of this equation gives:

$$W(\varphi) = P\cos\varphi + Q\sin\varphi \quad ,$$

where $P = const$ and $Q = const$ .

Then expressions (5.18a) and (5.17a) accordingly accept a kind:

$$v = \frac{P\cos\vartheta\cos\varphi + V\sin\vartheta + Q\cos\vartheta\sin\varphi}{r} \quad , \tag{5.18b}$$

$$w = \frac{\csc\vartheta(Q\cos\varphi - P\sin\varphi)}{r} \quad . \tag{5.17b}$$

As a result the equation (5.12b) gets a form:

$$\frac{V\cos\vartheta - \sin\vartheta(P\cos\varphi + Q\sin\varphi) + u}{r} = 0 \quad , \tag{5.12c}$$

whence

$$u = -V\cos\vartheta + \sin\vartheta(P\cos\varphi + Q\sin\varphi) \quad . \tag{5.19}$$

Substitution of values of functions $v$, $w$ и $u$ accordingly from formulas (5.18b), (5.17b) and (5.19) in the equations (5.11) and (5.14) transforms the last two into the equations:

$$\begin{aligned}&\frac{1}{h}(-V\sin\vartheta - \cos\vartheta(P\cos\varphi + Q\sin\varphi) + h(V\sin\vartheta + \cos\vartheta(P\cos\varphi + Q\sin\varphi)) + \\ &+\sqrt{hl}\,\tau_\vartheta) = 0\end{aligned} \quad , \tag{5.11a}$$

$$\sin\vartheta(Q\cos\varphi - P\sin\varphi) + \frac{\sin\vartheta(-Q\cos\varphi + P\sin\varphi)}{h} + \frac{\sqrt{l}\,\tau_\varphi}{\sqrt{h}} = 0 \quad . \tag{5.14a}$$

Integration of the equation (5.11a) gives

$$\tau = -\frac{(-1+h)(V\cos\vartheta + \sin\vartheta(P\cos\varphi + Q\sin\varphi))}{\sqrt{hl}} + F(t,r,\varphi) \quad .$$

Integration of the equation (5.14a) gives

$$\tau = -\frac{(-1+h)\sin\vartheta(P\cos\varphi + Q\sin\varphi)}{\sqrt{hl}} + \Phi(t,r,\vartheta) \quad .$$

Comparison of last two expressions gives

$$\tau = -\frac{(-1+h)(V\cos\vartheta + \sin\vartheta(P\cos\varphi + Q\sin\varphi))}{\sqrt{hl}} + F(t,r) \quad . \tag{5.20}$$

Now if to enter a designation

$$T = V\cos\vartheta - \sin\vartheta(P\cos\varphi + Q\sin\varphi) \quad ,$$

it is easy to be convinced that the equations (5.1), (5.2), (5.5) and (5.8) get forms:

$$\frac{-T\,l\,h' - T\,h\,l\,h' + T\sqrt{h\,l}\,\dot{h} - h^{\frac{3}{2}}\sqrt{l}\,(2l\,F' + F\,l' - 2T\,\dot{h}) + h^2(-T\,l' + 2l\,\dot{F} + F\,\dot{l})}{2h^2\,l^{\frac{3}{2}}} = 0 \quad , \tag{5.1a}$$

$$\frac{T\,h^{\frac{3}{2}}\,l' + T\sqrt{l}\,\dot{h} + h\sqrt{l}\,(F\,l' - 2T\,\dot{h}) + \sqrt{h}\,(-T\,l' + l\,(T\,h' - F\,\dot{h}))}{2h^2\,l} = 0 \quad , \tag{5.2a}$$

$$\frac{-T\sqrt{h}\,l' + F\,h\sqrt{l}\,l' + T\sqrt{l}\,\dot{h} + h^{\frac{3}{2}}(2l\,\dot{F} + F\,\dot{l})}{2h^2\sqrt{l}} = 0 \quad , \tag{5.5a}$$

$$\frac{T\sqrt{l}\,h' + h^{\frac{3}{2}}(2l\,F' + F\,l') - T\sqrt{h}\,\dot{h} + F\,h\sqrt{l}\,\dot{h}}{2h^2\sqrt{l}} = 0 \quad . \tag{5.8a}$$

All other equations of system (5.1-16) are satisfied.
Now we notice that the part of the members standing in the left parts of the equations of system (5.1a), (5.2a), (5.5a), (5.8a) does not depend on $T = T(\vartheta, \varphi)$ (i.e. depends on $t$ и $r$ only). Other part of members depends on $T$ linearly. It is obvious that the sums of the members depending and not dependent on $T$, should be equal zero separately. I.e. each of the equations of the above-stated system breaks up to two. Collecting the members which are not dependent on $T$, and the members which are the factors before $T$, we obtain system of eight equations:

$$\frac{-l\,h' - h\,l\,h' + \sqrt{h\,l}\,\dot{h} + 2h^{\frac{3}{2}}\sqrt{l}\;\dot{h} - h^2\,l'}{2h^2\,l^{\frac{3}{2}}} = 0 \tag{5.1b}$$

$$\frac{-h^{\frac{3}{2}}\sqrt{l}\,(2l\,F' + F\,l') + h^2(+2l\,\dot{F} + F\,\dot{l})}{2h^2\,l^{\frac{3}{2}}} = 0 \tag{5.1c}$$

$$\frac{h^{\frac{3}{2}}\,l' + \sqrt{l}\,\dot{h} - 2h\sqrt{l}\,\dot{h} + \sqrt{h}\,(-l' + l\,h')}{2h^2\,l} = 0 \tag{5.2b}$$

$$\frac{F\,(\sqrt{h}\,l' - \sqrt{l}\,\dot{h})}{2h^{\frac{3}{2}}\sqrt{l}} = 0 \tag{5.2c}$$

$$\frac{-\sqrt{h}\,l' + \sqrt{l}\;\dot{h}}{2h^2\sqrt{l}} = 0 \tag{5.5b}$$

$$\frac{F\,h\sqrt{l}\,l' + h^{\frac{3}{2}}(2l\,\dot{F} + F\,\dot{l})}{2h^2\sqrt{l}} = 0 \tag{5.5c}$$

$$\frac{\sqrt{l}\,h' - \sqrt{h}\,\dot{h}}{2h^2\sqrt{l}} = 0 \tag{5.8b}$$

$$\frac{\sqrt{h}\,(2l\,F' + F\,l') + F\sqrt{l}\,\dot{h}}{2h\sqrt{l}} = 0 \tag{5.8c}$$

Last system of the differential equations can be considered as linear homogeneous system for variables $\dot{h}, h', \dot{l}, l', \dot{F}, F'$. This system is untrivially joint. In particular, it has such two decisions:

$$h' = \frac{h\,l'}{l} \; ,$$
$$F' = -\frac{F\,l'}{l} \; ,$$

integrating which, we obtain:

$$h(t,r) = l(t,r)\, X(t) \, ,$$
$$F(t,r) = \frac{Y(t)}{l(t,r)} \; .$$

Substitution of these values in the equations (5.1b) and (5.1c) transforms the last two ones into the equations:

$$l(t,r)\,\dot{X}(t) - \sqrt{X(t)}\, l'(t,r) + X(t)\,\dot{l}(t,r) = 0 \; , \tag{5.1d}$$

$$2\,l(t,r)\sqrt{X(t)}\,\dot{Y}(t) + Y(t)\,l'(t,r) - \sqrt{X(t)}\,Y(t)\,\dot{l}(t,r) = 0 \, . \tag{5.1e}$$

During the deriving of these two equations we for the sake of simplification of a kind of the equation in the left parts have cleaned factors which cannot be zero or give trivial decisions.
If now from the equation (5.1d) we find $\dot{l}(t,r)$ and substitute this value in the equation (5.1e) the last one becomes:

$$\frac{l(t,r)(Y(t)\,\dot{X}(t) + 2\,X(t)\,\dot{Y}(t))}{\sqrt{X(t)}} = 0 \; .$$

This equation is easily integrated:

$$Y(t) = \frac{S}{\sqrt{X(t)}} \; .$$

Then function $F$ gets a form:

$$F(t,r) = \frac{S}{l(t,r)\sqrt{X(t)}} \; .$$

Substitution of last obtained values of functions $F$ and $h$ in the equation (5.8c) (with the simplification of a kind of the equation described above) gives:

$$l(t,r)\,\dot{X}(t) - \sqrt{X(t)}\, l'(t,r) + X(t)\,\dot{l}(t,r) = 0 \; .$$

Integration of the last equation for function $l$ gives:

$$l(t,r)=\frac{K(r+\int\frac{dt}{\sqrt{X(t)}})}{X(t)}\ .$$

If now we enter a designation $Z(t)=\int\frac{dt}{\sqrt{X(t)}}$, then, obviously $X(t)=\frac{1}{\dot{Z}^2(t)}$. Then

$$l(t,r)=K(r+Z(t))\dot{Z}^2(t)\ ,$$

$$h(t,r)=K(r+Z(t))\ ,$$

$$F(t,r)=\frac{S}{K(r+Z(t))\dot{Z}(t)}\ .$$

So, the central-symmetric metric is:

$$ds^2=-K(r+Z(t))dr^2-r^2(d\vartheta^2+\sin^2\vartheta d\varphi^2)+K(r+Z(t))\dot{Z}^2(t)dt^2\ ;$$

the function differentiable from the left is:

$$\vec{u}=(\frac{S+(-1+K(r+Z(t)))(V\cos\vartheta-\sin\vartheta(P\cos\varphi+Q\sin\varphi))}{K(r+Z(t))\dot{Z}(t)},\\ -V\cos\vartheta+\sin\vartheta(P\cos\varphi+Q\sin\varphi),\\ \frac{V\sin\vartheta+\cos\vartheta(P\cos\varphi+Q\sin\varphi)}{r},\\ \frac{Q\cos\varphi-P\sin\varphi}{r\sin\vartheta})\ .$$

Here $P,Q,V,S$ are arbitrary real constants, $Z(t)$ is an arbitrary function of time, $K(r+Z(t))$ is an arbitrary function of its argument.

## 6. The Consequences from the Geodetic Line Equations.

Let us copy out the equations ($4.8^1$) and (4.5) because they are very important for given below:

$$\frac{dx^0}{d\tau}=\vartheta^0\\ \frac{dx^1}{d\tau}=\vartheta^1\\ \frac{dx^2}{d\tau}=\vartheta^2\\ \frac{dx^3}{d\tau}=\vartheta^3\ ,\qquad (6.1)$$

$$\vartheta^i = \sum_j \alpha^j a^i_j \quad . \tag{6.2}$$

Don't forget please that the components of triangular transformation matrix $a^i_j$ can be found as functions of the components of metric tensor $g_{ij}$ .

If we demand the curve of zero length (see the parities (4.11)) (6.1) to be at the same time the geodetic line then the equations

$$\frac{d^2x^i}{d\tau^2} + \sum_{j,k} \Gamma^i_{jk} \frac{dx^j}{d\tau}\frac{dx^k}{d\tau} = 0$$

are to be satisfied together with the equations (6.1).

Let us put the parities (6.1) into these equations. We get the following chains of parities.

$$\begin{aligned}
\frac{d^2x^i}{d\tau^2} &= \frac{d\vartheta^i}{d\tau} = \sum_j \alpha^j \frac{da^i_j}{d\tau} = \sum_j \alpha^j \sum_{k,l,m} \frac{\partial a^i_j}{\partial g_{kl}} \frac{\partial g_{kl}}{\partial x^m} \frac{dx^m}{d\tau} = \\
&= \sum_{j,k,l,m} \alpha^j \frac{\partial a^i_j}{\partial g_{kl}} \frac{\partial g_{kl}}{\partial x^m} \vartheta^m = \sum_{j,k,l,m} \alpha^j \frac{\partial a^i_j}{\partial g_{kl}} \frac{\partial g_{kl}}{\partial x^m} \sum_n \alpha^n a^m_n = \quad . \\
&= \sum_{j,k,l,m,n} \alpha^j \alpha^n \frac{\partial a^i_j}{\partial g_{kl}} a^m_n \frac{\partial g_{kl}}{\partial x^m} = \sum_{j,k,l,m,n} \alpha^j \alpha^n \frac{\partial a^i_j}{\partial g_{kl}} a^m_n h_{klm}
\end{aligned} \tag{6.3}$$

In the last equality the designation $h_{klm} = \frac{\partial g_{kl}}{\partial x^m}$ is introduced.

$$\begin{aligned}
\sum_{j,k} \Gamma^i_{jk} \frac{dx^j}{d\tau}\frac{dx^k}{d\tau} &= \sum_{j,k} \Gamma^i_{jk} \vartheta^j \vartheta^k = \sum_{j,k} \Gamma^i_{jk} \sum_l \alpha^l a^j_l \sum_m \alpha^m a^k_m = \\
&= \sum_{j,k,l,m} \Gamma^i_{jk} \alpha^l \alpha^m a^j_l a^k_m = \sum_{j,k,l,n} \alpha^j \alpha^n \Gamma^i_{lk} a^l_j a^k_n =
\end{aligned} \tag{6.4}$$

(in the last equality the sum indices are changed: $j \to l$, $l \to j$, $m \to n$ .)

$$\begin{aligned}
&= \sum_{j,k,l,n} \alpha^j \alpha^n \Big(\frac{1}{2} \sum_m g^{im} \Big(\frac{\partial g_{ml}}{\partial x^k} + \frac{\partial g_{mk}}{\partial x^l} - \frac{\partial g_{lk}}{\partial x^m}\Big)\Big) a^l_j a^k_n = \\
&= \sum_{j,k,l,m,n} \alpha^j \alpha^n \frac{1}{2} g^{im} (h_{mlk} + h_{mkl} - h_{lkm}) a^l_j a^k_n
\end{aligned} \quad . \tag{6.4}$$

Adding the first and the last parts of chains (6.3) and (6.4) correspondingly to each other gives:

$$\frac{d^2x^i}{d\tau^2}+\sum_{j,k}\Gamma^i_{jk}\frac{dx^j}{d\tau}\frac{dx^k}{d\tau}=\sum_{j,n}\alpha^j\alpha^n(\sum_{k,l,m}\frac{\partial a^i_j}{\partial g_{kl}}a^m_n h_{klm}+\sum_{k,l,m}\frac{1}{2}g^{im}a^l_j a^k_n h_{mlk}+$$
$$+\sum_{k,l,m}\frac{1}{2}g^{im}a^l_j a^k_n h_{mkl}-\sum_{k,l,m}\frac{1}{2}g^{im}a^l_j a^k_n h_{lkm}).$$

We change in the second item in round brackets the sum indices: $k\to m,\quad m\to k$ , in the third: $k\to l,\quad l\to m,\quad m\to k$ , and in the forth: $k\to l,\quad l\to k$ . We obtain:

$$\frac{d^2x^i}{d\tau^2}+\sum_{j,k}\Gamma^i_{jk}\frac{dx^j}{d\tau}\frac{dx^k}{d\tau}=\sum_{j,n}\alpha^j\alpha^n\sum_{k,l,m}(\frac{\partial a^i_j}{\partial g_{kl}}a^m_n+\frac{1}{2}g^{ik}a^l_j a^m_n+$$
$$+\frac{1}{2}g^{ik}a^m_j a^l_n-\frac{1}{2}g^{im}a^k_j a^l_n)h_{klm}= . \tag{6.5}$$
$$=\sum_{j,n}\alpha^j\alpha^n\sum_{k,l,m}A^{iklm}_{jn}h_{klm}=0$$

Here the expression in round brackets is designed as $A^{iklm}_{jn}$ . Then we design $B^i_{jn}=\sum_{k,l,m}A^{iklm}_{jn}h_{klm}$ . From the last equality (6.5) we get:

$$\sum_{j,n}\alpha^j\alpha^n B^i_{jn}=0 \ . \tag{6.6}$$

Or in detailed form

$$\begin{aligned}&\alpha^0\alpha^0B^i_{00}+\alpha^0\alpha^1B^i_{01}+\alpha^0\alpha^2B^i_{02}+\alpha^0\alpha^3B^i_{03}+\\&+\alpha^1\alpha^0B^i_{10}+\alpha^1\alpha^1B^i_{11}+\alpha^1\alpha^2B^i_{12}+\alpha^1\alpha^3B^i_{13}+\\&+\alpha^2\alpha^0B^i_{20}+\alpha^2\alpha^1B^i_{21}+\alpha^2\alpha^2B^i_{22}+\alpha^2\alpha^3B^i_{23}+\\&+\alpha^3\alpha^0B^i_{30}+\alpha^3\alpha^1B^i_{31}+\alpha^3\alpha^2B^i_{32}+\alpha^3\alpha^3B^i_{33}=0\end{aligned} . \tag{6.6$^1$}$$

It is clear that according to the fact that the constants $\alpha^i$, $i=1,2,3$ are arbitrary the conditions

$$\begin{aligned}&B^k_{ij}+B^k_{ji}=0,\quad i\neq j\\&B^k_{00}+B^k_{11}=0\\&B^k_{00}+B^k_{22}=0\\&B^k_{00}+B^k_{33}=0\end{aligned} \tag{6.7}$$

are to be carried out.
It is the system of quasilinear uniform first-order partial differential equations for metric tensor. The system contains 36 equations while as it is well known the quantity of independent components of metric tensor is 10. Thus the system is overdetermined [6, p.102]. Such systems can have no any solutions at all. But we shall not analyse is this system joint or not and we shall try to find at least any solution.

**The remark 1.** The demand the isotropic curve (6.1) along which the differentiable function is transfered in parallel to be geodetic line have rather unexpected consequence concluded in serious restrictions for the components of metric tensor.

The direct solving of system (6.7) is very difficult because of complexity of expression of the triangular transformation matrix $a_i^j$ components through the components of metric tensor (the presence of radicals). It is much more easy to represent metric tensor as a function of triangular transformation matrix (the absence of radicals).

Further we shall operate in standard way. Let us write the Euler equations for the function $F=\sum_{i,j} g_{ij}\frac{dx^i}{d\tau}\frac{dx^j}{d\tau}=\sum_{i,j} g_{ij}x^{i'}x^{j'}$ :

$$\Phi_k = F_{x^k} - \frac{d}{d\tau}F_{x^{k'}} = 0 .$$

Further

$$\begin{aligned} F_{x^k} &= \frac{\partial}{\partial x^k}\sum_{i,j} g_{ij}x^{i'}x^{j'} = \sum_{i,j}\frac{\partial g_{ij}}{\partial x^k}\vartheta^i\vartheta^j = \\ &= \sum_{i,j}\Big(\sum_{r,s}\frac{\partial g_{ij}}{\partial a_r^s}\frac{\partial a_r^s}{\partial x^k}\Big)\Big(\sum_m \alpha^m a_m^i\Big)\Big(\sum_n \alpha^n a_n^j\Big) = \quad , \\ &= \sum_{i,j,m,n,r,s} \alpha^m\alpha^n a_m^i a_n^j \frac{\partial g_{ij}}{\partial a_r^s} h_{rk}^s \end{aligned} \tag{6.8}$$

where $h_{rk}^s = \frac{\partial a_r^s}{\partial x^k}$ . Then

$$\begin{aligned} F_{x^{k'}} &= \frac{\partial}{\partial x^{k'}}\sum_{i,j} g_{ij}x^{i'}x^{j'} = 2\sum_i g_{ki}x^{i'} \;, \\ \frac{d}{d\tau}F_{x^{k'}} &= 2\sum_i\Big(\frac{dg_{ki}}{d\tau}x^{i'} + g_{ki}\frac{dx^{k'}}{d\tau}\Big) = \\ &= 2\sum_i\Big(\sum_{m,n,r}\frac{\partial g_{ki}}{\partial a_m^n}\frac{\partial a_m^n}{\partial x^r}x^{r'}\Big)x^{i'} + 2\sum_i g_{ki}\frac{dx^{k'}}{d\tau} = \\ &= 2\sum_{i,m,n,r}\frac{\partial g_{ki}}{\partial a_m^n}h_{mr}^n\Big(\sum_j \alpha^j a_j^r\Big)\Big(\sum_s \alpha^s a_s^i\Big) + 2\sum_i g_{ki}\Big(\sum_j \alpha^j\frac{da_j^i}{d\tau}\Big) = \\ &= 2\sum_{i,j,m,n,r,s}\alpha^j\alpha^s a_j^r a_s^i\frac{\partial g_{ki}}{\partial a_m^n}h_{mr}^n + 2\sum_{i,j} g_{ki}\alpha^j\Big(\sum_m\frac{\partial a_j^i}{\partial x^m}x^{m'}\Big) = \\ &= 2\sum_{i,j,m,n,r,s}\alpha^j\alpha^s a_j^r a_s^i\frac{\partial g_{ki}}{\partial a_m^n}h_{mr}^n + 2\sum_{i,j,m}\alpha^j g_{ki}h_{jm}^i\Big(\sum_n \alpha^n a_n^m\Big) = \\ &= 2\sum_{i,j,m,n,r,s}\alpha^j\alpha^s a_j^r a_s^i\frac{\partial g_{ki}}{\partial a_m^n}h_{mr}^n + 2\sum_{i,j,m,n}\alpha^j\alpha^n g_{ki}a_n^m h_{jm}^i \end{aligned} \quad . \tag{6.9}$$

Uniting (6.8) and (6.9) we obtain:

$$\Phi_k = \sum_{i,j,m,n,r,s} \alpha^m \alpha^n a_m^i a_n^j \frac{\partial g_{ij}}{\partial a_r^s} h_{rk}^s - 2 \sum_{i,j,m,n,r,s} \alpha^j \alpha^s a_j^r a_s^i \frac{\partial g_{ki}}{\partial a_m^n} h_{mr}^n - $$
$$- 2 \sum_{i,j,m,n} \alpha^j \alpha^n g_{ki} a_n^m h_{jm}^i .$$

In the second item we change the sum indices:
$i \to j, \quad j \to m, \quad m \to r, \quad n \to s, \quad r \to i, \quad s \to n$ ,
in the third: $i \to s, \quad j \to m, \quad m \to j$ , and we get:

$$\Phi_k = \sum_{m,n} \alpha^m \alpha^n [\sum_j a_n^j (\sum_{i,r,s} (a_m^i \frac{\partial g_{ij}}{\partial a_r^s} h_{rk}^s - 2a_m^i \frac{\partial g_{kj}}{\partial a_r^s} h_{ri}^s) - \sum_s 2 g_{ks} h_{mj}^s)] =$$
$$= \sum_{m,n,k} \alpha^m \alpha^n B_{mnk} . \quad (6.10)$$

In the last equality the expression in square brackets is designed by symbol $B_{mnk}$ .
As like as in the previous case the system have the form:

$$\begin{aligned} & B_{ijk} + B_{jik} = 0, \quad i \neq j \\ & B_{00k} + B_{11k} = 0 \\ & B_{00k} + B_{22k} = 0 \\ & B_{00k} + B_{33k} = 0 \end{aligned} . \quad (6.11)$$

It is as before the overdetermined system of 36 quasilinear uniform partial differential equations for 10 unknown components of triangular matrix $a_i^j$ .
But it can be considered as a system of 36 algebraic linear uniform equations for 40 unknown $h_{ij}^k$ . Let us solve it.
Equations $B_{110} + B_{220} = 0, B_{110} + B_{330} = 0, B_{110} + B_{440} = 0$ immediately give $h_{10}^1 = 0, h_{20}^2 = 0, h_{30}^3 = 0$ consequently.
The substitution of these values into the system leads to the fact that the equations $B_{210} + B_{120} = 0, B_{320} + B_{230} = 0$ give $h_{20}^1 = 0, h_{30}^2 = 0$ .
Let us substitute the obtained values into the system again. We can get that the equation $B_{310} + B_{130} = 0$ gives $h_{30}^1 = 0$ .
Further we shall operate similarly (in the sense of substitution of obtained values).
The left parts of three equations $B_{100} + B_{010} = 0, B_{101} + B_{011} = 0, B_{000} + B_{111} = 0$ differ from each other by factors only, that is why from these equations we can get the only one value $h_{ij}^k$ expressing it through the other, for example:

$$h_{10}^0 = \frac{a_1^0 h_{00}^0 + a_1^1 h_{01}^0}{a_0^0} . \quad (6.12)$$

Then the equations $B_{001} + B_{221} = 0, B_{001} + B_{331} = 0$ give $h_{21}^2 = 0, h_{31}^3 = 0$ .
The equation $B_{321} + B_{231} = 0$ gives $h_{31}^2 = 0$ .

Further the left parts of two equations $B_{211}+B_{121}=0$, $B_{212}+B_{122}=0$ differ from each other by factors only. They give

$$h_{21}^{1}=\frac{a_2^1 h_{11}^1 + a_2^2 h_{12}^1}{a_1^1}\ . \tag{6.13}$$

Again the left parts of three equations $B_{002}+B_{112}=0$, $B_{200}+B_{020}=0$, $B_{002}+B_{222}=0$ differ from each other by factors. From here we get

$$h_{20}^{0}=\frac{a_2^0 h_{00}^0 + a_2^1 h_{01}^0 + a_2^2 h_{02}^0}{a_0^0}\ . \tag{6.14}$$

The solving of two-equation system $B_{322}+B_{232}=0$, $B_{323}+B_{233}=0$ gives

$$h_{32}^{2}=\frac{a_3^2 h_{22}^2 + a_3^3 h_{23}^2}{a_2^2}\ , \tag{6.15}$$

$$h_{32}^{3}=0$$

And again the left parts of two equations $B_{102}+B_{012}=0$, $B_{201}+B_{021}=0$ differ from each other by factors. We obtain

$$h_{21}^{0}=\frac{a_1^1 a_2^0 h_{01}^0 - a_1^0 a_2^1 h_{01}^0 - a_1^0 a_2^2 h_{02}^0 + a_0^0 a_2^1 h_{11}^0 + a_0^0 a_2^2 h_{12}^0}{a_0^0 a_1^1}\ . \tag{6.16}$$

The left parts of three equations $B_{003}+B_{223}=0$, $B_{300}+B_{030}=0$, $B_{312}+B_{132}=0$ differ from each other by factors. From here we get

$$h_{30}^{0}=\frac{a_3^0 h_{00}^0 + a_3^1 h_{01}^0 + a_3^2 h_{02}^0 + a_3^3 h_{03}^0}{a_0^0}\ . \tag{6.17}$$

And again the left parts of two equations $B_{003}+B_{113}=0$, $B_{311}+B_{131}=0$ differ from each other by factors only. From here we can obtain

$$h_{31}^{1}=\frac{a_3^1 h_{11}^1 + a_3^2 h_{12}^1 + a_3^3 h_{13}^1}{a_1^1}\ . \tag{6.18}$$

The left parts of three equations $B_{213}+B_{123}=0$, $B_{312}+B_{132}=0$, $B_{313}+B_{133}=0$ differ from each other by factors only. We get

$$h_{32}^{1}=\frac{a_2^2 a_3^1 h_{12}^1 - a_2^1 a_3^2 h_{12}^1 - a_2^1 a_3^3 h_{13}^1 + a_1^1 a_3^2 h_{22}^1 + a_1^1 a_3^3 h_{23}^1}{a_1^1 a_2^2}\ . \tag{6.19}$$

From the equation $B_{103}+B_{013}=0$ we can find

$$h_{31}^0 = \frac{a_1^1 a_3^0 h_{01}^0 - a_1^0(a_3^1 h_{01}^0 + a_3^2 h_{02}^0 + a_3^3 h_{03}^0) + a_0^0(a_3^1 h_{11}^0 + a_3^2 h_{12}^0 + a_3^3 h_{13}^0)}{a_0^0 a_1^1} \quad . \tag{6.20}$$

The equation $B_{203} + B_{023} = 0$ gives

$$\begin{aligned} h_{32}^0 = {} & \frac{1}{a_0^0 a_1^1 a_2^2}(a_1^0(-a_2^2 a_3^1 h_{02}^0 + a_2^1 a_3^2 h_{02}^0 + a_2^1 a_3^3 h_{03}^0) + a_0^0(a_2^2 a_3^1 h_{12}^0 - a_2^1(a_3^2 h_{12}^0 + a_3^3 h_{13}^0)) + \\ & + a_1^1(a_2^2 a_3^0 h_{02}^0 - a_2^0(a_3^2 h_{02}^0 + a_3^3 h_{03}^0) + a_0^0(a_3^2 h_{22}^0 + a_3^3 h_{23}^0))) \end{aligned} \quad . \tag{6.21}$$

After this step the system of algebraic linear equations is completely solved.
The conditions $h_{30}^3 = 0, h_{31}^3 = 0, h_{32}^3 = 0$ give the dependence $a_3^3 = a_3^3(x^3)$ .
The conditions $h_{30}^2 = 0, h_{31}^2 = 0$ give the dependence $a_3^2 = a_3^2(x^2, x^3)$ .
The conditions $h_{20}^2 = 0, h_{21}^2 = 0$ give the dependence $a_2^2 = a_2^2(x^2, x^3)$ .
And the conditions $h_{30}^1 = 0, h_{20}^1 = 0, h_{10}^1 = 0$ give the dependences
$a_3^1 = a_3^1(x^1, x^2, x^3), a_2^1 = a_2^1(x^1, x^2, x^3), a_1^1 = a_1^1(x^1, x^2, x^3)$ consequently.
Differential equations (6.12-21) are to be solved.
The equation (6.21) have the solutions:

$$\begin{aligned} a_2^2 &= -\frac{1}{F^{1,0}(x^2, x^3)} \\ a_3^2 &= -\frac{a_3^3(x^3) F^{0,1}(x^2, x^3)}{F^{1,0}(x^2, x^3)} \end{aligned} \quad , \tag{6.22}$$

where $F(x^2, x^3)$ is an arbitrary function. Upper symbols of course mean the differentiation by the corresponding argument.
Don't forget that we are looking for any solution, that is why when during the integration of this equation additive to the main solution arbitrary functions appeared we made them equal to zero for the purpose of simplification.
The obtained values of the unknown functions we shall substitute into the other equations of the system immediately.
The solutions of the equation (6.18) are

$$\begin{aligned} a_1^1 &= -\frac{1}{H^{1,0,0}(x^1, x^2, x^3)} \\ a_2^1 &= \frac{H^{0,1,0}(x^1, x^2, x^3)}{F^{1,0}(x^2, x^3) H^{1,0,0}(x^1, x^2, x^3)} \end{aligned} \quad , \tag{6.23}$$

where $H(x^1, x^2, x^3)$ is an arbitrary function also.
The solution of the equation (6.19) is

$$a_3^1 = \frac{a_3^3(x^3)(-H^{0,0,1}(x^1, x^2, x^3) + \dfrac{F^{0,1}(x^2, x^3) H^{0,1,0}(x^1, x^2, x^3)}{F^{1,0}(x^2, x^3)})}{H^{1,0,0}(x^1, x^2, x^3)} \quad . \tag{6.24}$$

The solutions of the equation (6.12) are

$$a_0^0 = -\frac{1}{P^{1,0,0,0}(x^0,x^1,x^2,x^3)}$$
$$a_1^0 = \frac{P^{0,1,0,0}(x^0,x^1,x^2,x^3)}{H^{1,0,0}(x^1,x^2,x^3)P^{1,0,0,0}(x^0,x^1,x^2,x^3)} \quad , \tag{6.25}$$

where $P(x^0,x^1,x^2,x^3)$ - is an arbitrary function again.
The solution of the equation (6.13) is

$$a_2^0 = \frac{H^{1,0,0}(x^1,x^2,x^3)P^{0,0,1,0}(x^0,x^1,x^2,x^3) - H^{0,1,0}(x^1,x^2,x^3)P^{0,1,0,0}(x^0,x^1,x^2,x^3)}{F^{1,0}(x^2,x^3)H^{1,0,0}(x^1,x^2,x^3)P^{1,0,0,0}(x^0,x^1,x^2,x^3)} \quad . \tag{6.26}$$

The solution of the equation (6.15) is

$$\begin{aligned} a_3^0 = {} & (a_3^3(x^3)(H^{1,0,0}(x^1,x^2,x^3)(-F^{1,0}(x^2,x^3)P^{0,0,0,1}(x^0,x^1,x^2,x^3) + \\ & + F^{0,1}(x^2,x^3)P^{0,0,1,0}(x^0,x^1,x^2,x^3)) + \\ & + (F^{1,0}(x^2,x^3)H^{0,0,1}(x^1,x^2,x^3) - F^{0,1}(x^2,x^3)H^{0,1,0}(x^1,x^2,x^3)) \\ & P^{0,1,0,0}(x^0,x^1,x^2,x^3)))/ \\ & /(F^{1,0}(x^2,x^3)H^{1,0,0}(x^1,x^2,x^3)P^{1,0,0,0}(x^0,x^1,x^2,x^3)) \end{aligned} \quad . \tag{6.27}$$

The remained 4 equations are satisfied automatically. Thus one special solution for the triangular transformation matrix is found and can be represented by formulas (6.22-27). The function $a_3^3(x^3)$ is of course arbitrary also.
The solving of algebraic system of equations connecting triangular transformation matrix and metric tensor gives the values of the components of it:

$$\begin{aligned} g_{00} &= P^{1,0,0,0^2}(x^0,x^1,x^2,x^3) \\ g_{01} &= P^{0,1,0,0}(x^0,x^1,x^2,x^3)P^{1,0,0,0}(x^0,x^1,x^2,x^3) \\ g_{02} &= P^{0,0,1,0}(x^0,x^1,x^2,x^3)P^{1,0,0,0}(x^0,x^1,x^2,x^3) \\ g_{03} &= P^{0,0,0,1}(x^0,x^1,x^2,x^3)P^{1,0,0,0}(x^0,x^1,x^2,x^3) \\ g_{11} &= -H^{1,0,0^2}(x^1,x^2,x^3) + P^{0,1,0,0^2}(x^0,x^1,x^2,x^3) \\ g_{12} &= -H^{0,1,0}(x^1,x^2,x^3)H^{1,0,0}(x^1,x^2,x^3) + P^{0,0,1,0}(x^0,x^1,x^2,x^3)P^{0,1,0,0}(x^0,x^1,x^2,x^3) \quad . \\ g_{13} &= -H^{0,0,1}(x^1,x^2,x^3)H^{1,0,0}(x^1,x^2,x^3) + P^{0,0,0,1}(x^0,x^1,x^2,x^3)P^{0,1,0,0}(x^0,x^1,x^2,x^3) \\ g_{22} &= -F^{1,0^2}(x^2,x^3) - H^{0,1,0^2}(x^1,x^2,x^3) + P^{0,0,1,0^2}(x^0,x^1,x^2,x^3) \\ g_{23} &= -F^{0,1}(x^2,x^3)F^{1,0}(x^2,x^3) - H^{0,0,1}(x^1,x^2,x^3)H^{0,1,0}(x^1,x^2,x^3) + \\ & \quad + P^{0,0,0,1}(x^0,x^1,x^2,x^3)P^{0,0,1,0}(x^0,x^1,x^2,x^3) \\ g_{33} &= -\frac{1}{a_3^{3^2}(x^3)} - F^{0,1^2}(x^2,x^3) - H^{0,0,1^2}(x^1,x^2,x^3) + P^{0,0,0,1^2}(x^0,x^1,x^2,x^3) \end{aligned}$$

**The remark 2**. The obtained special solution don't correspond to the central-symmetric solution from the paragraph 5. Of course this fact has place because the special solution contains too small arbitrarity in functions. The general solution of the equations (6.12, 13,15, 18, 19, 21 ) contains another 13 arbitrary functions of different sets of arguments except four mentioned above.(This general solution we shall not introduce here.) I could not solve the equations (6.14, 16, 17, 20) in general because of their complexity. These equations must concretize the form of some of arbitrary functions. However rather simple analysis of obtained general solutions of the equations (6.12, 13, 15, 18, 19, 21) shows that some of them may correspond to central-symmetric case.

**The remark 3**. Frankly speaking the demand that the curves given by the first-order differential equations (6.1) must be at the same time geodetic lines does not have a relation to the algebra and analysis mentioned above and they are caused nothing but aesthetic view. It simply will not be fine if waves spreading in flat space along straight lines will spread in curved space not along geodetic lines. Thus this idea is some "incidental injection" into the stated theory.

**The remark 4**. In general if to demand that the curve given by equations

$\frac{dx^i}{d\tau} = f^i(x^1,...,x^n)$ would contain in the right side not simply functions of coordinates but functions of metric tensor (in its turn depending on coordinates)

$f^i(g_{00}(x^0,x^1,x^2,x^3),...,g_{ij}(x^0,x^1,x^2,x^3),...,g_{nn}(x^0,x^1,x^2,x^3))$ and would at the same time be a geodetic line it will be clear that this demand will restrict the form of metric tensor.

**The remark 5**. The joint (general) solution of systems (6.7) and (4.9) (or introduction of general solution of system (6.7) into the system (4.9), what is the same operation) exhausts all possible values of metric tensor and all possible values of differentiable function. Attracting additional considerations about the form of metric tensor (considerations of symmetry for example) will contract the list of possible solutions.

## 7. Some Additional Results.

**1**. We shall try to analyse do associative algebras with the special law of multiplication of an element and its conjugated one

$$\begin{aligned}
\vec{r}\vec{r}' &= (x^0\vec{e}_0 + x^1\vec{e}_1 + x^2\vec{e}_2 + x^3\vec{e}_3 + ...)(x^0\vec{e}_0 - x^1\vec{e}_1 - x^2\vec{e}_2 - x^3\vec{e}_3 - ...) = \\
&= (x^0)^2\vec{e}_0\vec{e}_0 + x^0x^1\vec{e}_1\vec{e}_0 + x^0x^2\vec{e}_2\vec{e}_0 + x^0x^3\vec{e}_3\vec{e}_0 + ... - \\
&- x^0x^1\vec{e}_0\vec{e}_1 - (x^1)^2\vec{e}_1\vec{e}_1 - x^1x^2\vec{e}_2\vec{e}_1 - x^1x^3\vec{e}_3\vec{e}_1 - ... - \\
&- x^0x^2\vec{e}_0\vec{e}_2 - x^1x^2\vec{e}_1\vec{e}_2 - (x^2)^2\vec{e}_2\vec{e}_2 - x^2x^3\vec{e}_3\vec{e}_2 - ... - \\
&- x^0x^3\vec{e}_0\vec{e}_3 - x^1x^3\vec{e}_1\vec{e}_3 - x^2x^3\vec{e}_2\vec{e}_3 - (x^3)^2\vec{e}_3\vec{e}_3 - ... - \\
&- ... = \\
&= ((x^0)^2 - (x^1)^2 - (x^2)^2 - (x^3)^2 - ...)\vec{f}_0
\end{aligned} \tag{7.1}$$

in spaces of dimension $n \neq 4$ exist.

The requirement (7.1) imposes the following conditions on factors $a_{ijk}$ of decomposition of product of elements $\vec{e}_i\vec{e}_j = a_{ij0}\vec{e}_0 + a_{ij1}\vec{e}_1 + ...$ by basis $\{\vec{e}_i\}$ of space:

$\vec{e}_i\vec{e}_0 = a_{i00}\vec{e}_0 + a_{i01}\vec{e}_1 + ... = \vec{e}_0\vec{e}_i = a_{0i0}\vec{e}_0 + a_{0i1}\vec{e}_1 + ... \qquad \forall\, 1 \le i \le n$ or

$$a_{i0j} = a_{0ij} \qquad \forall\, 1 \le i \le n, \quad 0 \le j \le n \qquad ; \qquad (7.2)$$

$$\vec{e}_0\vec{e}_0 = a_{000}\vec{e}_0 + a_{001}\vec{e}_1 + ... = \vec{e}_i\vec{e}_i = a_{ii0}\vec{e}_0 + a_{ii1}\vec{e}_1 + ... \qquad \forall\, 1 \le i \le n \qquad \text{or}$$

$$a_{00j} = a_{iij} \qquad \forall\, 1 \le i \le n, \quad 0 \le j \le n \qquad ; \qquad (7.3)$$

$$\vec{e}_i\vec{e}_j = a_{ij0}\vec{e}_0 + a_{ij1}\vec{e}_1 + ... = -\vec{e}_j\vec{e}_i = -a_{ji0}\vec{e}_0 - a_{ji1}\vec{e}_1 - ... \qquad \forall\, 1 \le i < j \le n \qquad \text{or}$$

$$a_{ijk} = -a_{jik} \qquad \forall\, 1 \le i < j \le n, \quad 0 \le k \le n \quad . \qquad (7.4)$$

Requirement of associativity of multiplication imposes the condition

$$\vec{e}_i(\vec{e}_j\vec{e}_k) = \vec{e}_i\sum_{l=0}^{n} a_{jkl}\vec{e}_l = \sum_{l=0}^{n} a_{jkl}\vec{e}_i\vec{e}_l = \sum_{l=0}^{n} a_{jkl}\sum_{m=0}^{n} a_{ilm}\vec{e}_m = \sum_{m=0}^{n}(\sum_{l=0}^{n} a_{jkl}a_{ilm})\vec{e}_m =$$

$$= (\vec{e}_i\vec{e}_j)\vec{e}_k = (\sum_{l=0}^{n} a_{ijl}\vec{e}_l)\vec{e}_k = \sum_{l=0}^{n} a_{ijl}\vec{e}_l\vec{e}_k = \sum_{l=0}^{n} a_{ijl}\sum_{m=0}^{n} a_{lkm}\vec{e}_m = \sum_{m=0}^{n}(\sum_{l=0}^{n} a_{ijl}a_{lkm})\vec{e}_m \qquad \text{or}$$

$$\sum_{l=0}^{n}(a_{jkl}a_{ilm} - a_{ijl}a_{lkm}) = 0 \qquad \forall\, 0 \le i, j, k, m \le n \quad . \qquad (7.5)$$

on factors $a_{ijk}$ .

The solving of system (7.2) – (7.5) gives the following rules of multiplication of basic vectors when $n = 2$:

$$\vec{e}_0\vec{e}_0 = \vec{e}_0\vec{e}_1 = \vec{e}_1\vec{e}_0 = \vec{e}_1\vec{e}_1 = \vec{e}_0 + \vec{e}_1 \qquad ;$$

when $n = 3$ - trivial solutions only:

$$\vec{e}_i\vec{e}_j = 0 \qquad \forall\, 0 \le i, j \le 3 \qquad ;$$

when $n = 5$:

$$\vec{e}_0\vec{e}_0 = \vec{e}_1\vec{e}_1 = \vec{e}_2\vec{e}_2 = \vec{e}_3\vec{e}_3 = \vec{e}_4\vec{e}_4 = a\vec{R}$$
$$\vec{e}_0\vec{e}_1 = \vec{e}_1\vec{e}_0 = -\alpha_1 a\vec{R}$$
$$\vec{e}_0\vec{e}_2 = \vec{e}_2\vec{e}_0 = -\alpha_2 a\vec{R}$$
$$\vec{e}_0\vec{e}_3 = \vec{e}_3\vec{e}_0 = -\alpha_3 a\vec{R}$$
$$\vec{e}_0\vec{e}_4 = \vec{e}_4\vec{e}_0 = -\alpha_4 a\vec{R}$$
$$\vec{e}_i\vec{e}_j = 0 \qquad \forall\, 1 \le i, j \le 5; \quad i \ne j$$

where $a = const$ is an arbitrary real constant, $\alpha_i$ are real constants which obey the condition $(\alpha_1)^2 + (\alpha_2)^2 + (\alpha_3)^2 + (\alpha_4)^2 = 1$; $\vec{R} = \vec{e}_0 + \alpha_1\vec{e}_1 + \alpha_2\vec{e}_2 + \alpha_3\vec{e}_3 + \alpha_4\vec{e}_4$. (Here constants $\alpha$ have lower indices because it is not necessary to examine tensors).

Research of algebras of this kind in spaces of higher dimension is interfaced to the big computing difficulties.

**2**. If to speak about geometrical or other sense of constants $a$ and $c$, entering into algebraic parities (1.11), here, unlike constants $\alpha_i$, clearness is not present (here because we do not mean any tensors parameters $\alpha_i$ have lower indices). We take for a basis a special kind of algebra (1.11), namely (3.3), and we shall applicate to basic vectors one-dimensional Lorentz's transformation

$$\vec{n}_0 = \frac{1}{\sqrt{1-\beta^2}}\vec{k}_0 + \frac{\beta}{\sqrt{1-\beta^2}}\vec{k}_1$$
$$\vec{n}_1 = \frac{\beta}{\sqrt{1-\beta^2}}\vec{k}_0 + \frac{1}{\sqrt{1-\beta^2}}\vec{k}_1 \quad .$$
$$\vec{n}_2 = \vec{k}_2$$
$$\vec{n}_3 = \vec{k}_3$$

Then the algebra of basic vectors $\vec{n}_i$ became:

$$\vec{n}_0\vec{n}_0 = \frac{a(-1+\beta)^2}{(1+\beta)\sqrt{1-\beta^2}}(\vec{n}_0+\vec{n}_1)$$
$$\vec{n}_0\vec{n}_1 = -\frac{a(-1+\beta)^2}{(1+\beta)\sqrt{1-\beta^2}}(\vec{n}_0+\vec{n}_1)$$
$$\vec{n}_0\vec{n}_2 = 0$$
$$\vec{n}_0\vec{n}_3 = 0$$
$$\vec{n}_1\vec{n}_0 = -\frac{a(-1+\beta)^2}{(1+\beta)\sqrt{1-\beta^2}}(\vec{n}_0+\vec{n}_1)$$
$$\vec{n}_1\vec{n}_1 = \frac{a(-1+\beta)^2}{(1+\beta)\sqrt{1-\beta^2}}(\vec{n}_0+\vec{n}_1)$$
$$\vec{n}_1\vec{n}_2 = 0$$
$$\vec{n}_1\vec{n}_3 = 0$$
$$\vec{n}_2\vec{n}_0 = 0$$
$$\vec{n}_2\vec{n}_1 = 0$$
$$\vec{n}_2\vec{n}_2 = -\frac{a(-1+\beta)}{\sqrt{1-\beta^2}}(\vec{n}_0+\vec{n}_1)$$
$$\vec{n}_2\vec{n}_3 = -\frac{c(-1+\beta)}{\sqrt{1-\beta^2}}(\vec{n}_0+\vec{n}_1)$$
$$\vec{n}_3\vec{n}_0 = 0 \quad .$$
$$\vec{n}_3\vec{n}_1 = 0$$
$$\vec{n}_3\vec{n}_2 = \frac{c(-1+\beta)}{\sqrt{1-\beta^2}}(\vec{n}_0+\vec{n}_1)$$
$$\vec{n}_3\vec{n}_3 = -\frac{a(-1+\beta)}{\sqrt{1-\beta^2}}(\vec{n}_0+\vec{n}_1)$$

In the right part of last parities before the vector $(\vec{n}_0 + \vec{n}_1)$ with a sign "plus" or "minus" is one of three multipliers $\frac{a(-1+\beta)^2}{(1+\beta)\sqrt{1-\beta^2}}$, $-\frac{a(-1+\beta)}{\sqrt{1-\beta^2}}$ or $\frac{c(-1+\beta)}{\sqrt{1-\beta^2}}$. No one of them at any $\beta$ can become equal $0$. But it is possible to choose such value of $\beta$ that, for example, the second of them became equal to any beforehand real number, say, to unity. Thus $\beta = \frac{-1+a^2}{1+a^2}$. It is obvious that $\forall -\infty < a < \infty$ inequalities $-1 < \beta < 1$ are carried out as, in general, it must be. At such value of parameter $\beta$ the first of the above-stated multipliers becomes equal to $\frac{1}{a^2}$, and the third – equal to $-\frac{c}{a}$. The sense of the specified constants $a$ and $c$ remains not clear. Besides it is necessary to notice that these constants do not enter into analogues of conditions of Cauchy-Riemann (2.2).

**3**. The question of what algebraic operations over functions give as a result differentiable functions, is much more complicated, than usually, and demands additional detailed research. For example, it is obvious that the sum of two differentiable functions is the differentiable function. However, this condition is not necessary.
Even more difficult is a situation with product of functions. It is clear that product of differentiable and any other function gives as a result a zero. Therefore differentiable and untrivial can be the product of not differentiable functions only.

**4**. System of equations (2.2) for spherically-symmetric case (i.e. when $x_0 = t, x_1 = \rho, x_2 = \varphi, x_3 = \vartheta$) in a case when $\alpha_1 = \alpha_\rho = 1$, seemingly, has no solutions. I.e. I honesty tried to find them, but I could not. Certainly, it at all does not mean that solutions do not exist. However, I think that solutions do not exist actually, and it is symptomatic. I shall leave comments for the Conclusion.
It is necessary to notice that cylindrically-symmetric solution in a case when $\alpha_3 = \alpha_z = 1$, can be found without effort.

## Conclusion.

My deep belief is that the world should submit to the field equations. From this point of view the equations of Einstein, and the equations of Maxwell as well, are the approximate description of the reality as in the right parts of these equations there are not field variables.

Topological ideology on which as I understand the theory of superstrings is based, and together with it the idea of not continuous space-time as similarly its methods are not related to me. In my opinion, this ideology has appeared as a result of despair of the fact that any equations of fundamental physics (it is not important, whether quantum theory, or General Theory of Relativity) in the solutions of a central-symmetric problem contain singularity in the centre. The present article (of course not pretending to be right in sense of physics) gives hope that while using not geometrical only, but also algebraic and analytical reasons, these difficulties will be managed to avoid.

I distinctly realize that in the given article instead of recognized principle of the least action (and together with it Hamilton's formalism) other mathematical formalism (probably the first is a consequence of the second or vice versa) is offered. It is, of course, very impudent. However, let's discuss the following circumstances. There are two types of field that are present in the analogues of conditions of Cauchy-Riemann (4.9): gravitational (by definition) and some other field. What is this field? I.e., whether it has any real correspondence? Solutions (concerning this field) of the analogues of conditions of Cauchy-Riemann in pseudo-Riemannian space are the waves extending with speed "unity" along zero geodetic lines. As they say, whose ears are sticking out from a hat?

Generally speaking the fundamental equations of physics are invariant in relation to time inversion (and of course charge and parity). However, time is irreversible. (Statistical reasons are of course very strong, but as it seems to me, are not quite satisfactory.) Maybe this fact indicates the chronic lacks of the basic fundamental equations of physics? Anyway, the theory stated here (certainly because there are equations of the first, instead of the second, order) contains as a kind of differentiable solutions "direct" waves only, not "reversed".

Further, if to speak about central-symmetric stationary metric, the solutions of Schwarzschild, Reissner-Nordstrom and so forth, and Newton's and Coulomb's laws as their approximations, simply should not exist as the solutions of the fundamental equations. Photons and "gravitons" only, i.e. wave solutions. Either we recognize interaction in a wave mode, spreading with final speed, or we recognize the presence of forces of type $\frac{1}{r^2}$ too. By the way, how does in this sense the recognition of collapsar existence in the spirit of the Schwarzschild's solution look? Nothing material can overcome event horizon, even photons. And what about gravitons? Or they cannot bear the information (by the way, this argument was never clear to me)?

Or the substitution of potential of type $\frac{1}{r}$ in the equation of Schroedinger gives as the solutions magically the waves for the transition levels of electrons in atom of hydrogen. We substitute a field extending infinitely quickly, and we obtain a field extending with final speed?

In my opinion, interactions are possible with final speed only, i.e. the fundamental equations of physics should have wave solutions only. Forces of type $\frac{1}{r^2}$ arise because of probability to absorb a photon (or "graviton") on distance $r$ (maybe, big enough) from a source. Thus, Einstein's and Maxwell's equations should correspond with more exact equations approximately in the same way as the thermodynamics equations correspond with the equations of statistical physics.

However, all these questions are the questions to the standard theory.

Further, absence of possibility in the stated theory to define components of derivative of differentiable function strongly resembles the principle of uncertainty of Heisenberg in its extreme formulation: "if we know everything about coordinates, we know nothing about impulse, and vice versa". This question demands the further research. Probably algebra stated here needs some modification. I shall explain my thought. Here "the one-particle theory" is stated, so to say. I.e. the matter is that product of basic vectors gives defined to within several constants, but defined (!) for the given algebra and space result. It seems that we speak about free fields. If we want to obtain interacting fields algebra (and analysis) stated here and depending on the set of constants $a, c, \alpha_1, \alpha_2, \alpha_3$ is not enough.

Absence of spherically-symmetric solutions of analogues of the equations of Cauchy-Riemann in pseudoeuclidean flat space (I really think that they do not exist) speaks, in my opinion, about such a fact: in flat space flat waves are only possible (photons?). Presence of exact (to within several constants and arbitrary functions) solutions of a central-symmetric problem in pseudo-Riemannian space, probably, testifies to existence of particle-like (Yu. S. Vladimirov) solutions of "fundamental" (we shall hope) equations, namely, massive particles. This question also demands the further studying.

I think that algebras of the type similar to the above-stated in spaces of the higher dimensions should possess properties, much more "poor", than in space of dimension "four". Accordingly, analytical functions should possess more "poor" properties. It is one more argument against the idea that our space has dimension more than "three".

**References.**